\documentclass[a4paper,fleqn]{cas-dc}

\usepackage[numbers,sort&compress]{natbib}
\usepackage{subcaption}
\usepackage{enumitem}
\usepackage{amsmath}
\def\tsc#1{\csdef{#1}{\textsc{\lowercase{#1}}\xspace}}
\tsc{WGM}
\tsc{QE}
\usepackage{algorithm}
\usepackage{algpseudocode}
\usepackage{booktabs}
\usepackage{mdframed}
\usepackage{multicol}
\usepackage{amsmath}
\usepackage{amssymb}
\usepackage{xcolor}
\definecolor{myred}{RGB}{180,40,40}
\definecolor{mypurple}{RGB}{120,60,180}
\definecolor{myblue}{RGB}{40,70,150}
\definecolor{negcol}{RGB}{163, 45, 45}   
\definecolor{poscol}{RGB}{15, 110, 86}   
\usepackage{siunitx}
\title[mode = title]{Cost-effective design of grid-connected community microgrids}

\begin{document}




\author[1]{Moslem Uddin}[type=editor,
auid=000,bioid=1,]
\cormark[1] 
\ead{moslem.uddin.bd@gmail.com} 

\credit{Conceptualization of this study, Methodology, Software,  Drafting}

\author[2]{Huadong Mo}[type=editor,
auid=000,bioid=1,]
\ead{huadong.mo@unsw.edu.au} 

\credit{Supervision, Knowledge, Review \& editing}

\author[3]{Daoyi Dong}[type=editor,
auid=000,bioid=1,]
\ead{Daoyi.Dong@uts.edu.au} 

\credit{Supervision, Knowledge, Review \& editing}

%

\address[1]{School of Engineering Technology, The University of New South Wales, Canberra, ACT 2610, Australia}
\address[2]{School of Systems and Computing, The University of New South Wales, Canberra, ACT 2610, Australia}
\address[3]{Australian AI Institute, FEIT, University of Technology Sydney, Sydney, NSW 2007, Australia}

\begin{abstract}[S U M M A R Y]
This study aims to develop a cost-effective microgrid (MG) design that optimally balances the economic feasibility, reliability, efficiency, and environmental impact in a grid-tied community MG. A multi-objective optimization framework is first employed to generate feasible MG configurations considering economic, reliability, efficiency, and environmental objectives. Subsequently, a preference-based deep reinforcement learning (DRL) framework is utilized to evaluate and select preferred configurations using a scalarized reward function. This combined approach enables systematic exploration of trade-offs among conflicting objectives and supports informed decision-making for community MG planning.
Sensitivity analyses are conducted to evaluate the system performance under varying load demand and renewable energy fluctuations. Beside, an economic sensitivity assessment examines the impact of electricity prices and capital costs on the levelized cost of energy (LCOE).
The proposed MG configuration achieves high reliability, satisfying 100\% of the load, even under adverse weather conditions. The proposed framework attains an efficiency of 91.99\% while maintaining a carbon footprint of 302,747 kg/year, which is approximately 95\% lower than the annual emissions associated with a conventional grid-supplied energy system. The economic analysis indicates a net present cost of \$4.83M with a competitive LCOE  of \$0.208/kWh. In addition, the operation cost is \$201,473 per year with a capital investment of \$1.42M, rendering it a financially viable alternative to conventional grid-dependent systems.
This work can be valuable in identifying effective solutions for supplying reliable and cost-effective
power to regional and remote areas.
\end{abstract}


\maketitle

\begin{figure*}[!tp]  	
	\centering
	\begin{mdframed}[linewidth=1pt]
		\begin{multicols}{2}
			\textbf{Nomenclature}
			\begin{description}
				\item[$ \mathcal{C} $] Cost metric of the MG
				\item[$ \xi $] Emission metric of the MG
				\item[$ \eta $] Efficiency metric of the MG
				\item[$ \mathcal{R} $] Reliability metric of the MG	
				\item [MO-GA] Multi-objective GA
				\item [MO-OPT] Multi-Objective Optimization
				\item [ILP] Integer Linear Programming
				\item [MCS] Monte Carlo Simulation
				\item [PESA-II] Pareto Envelope-based Selection Algorithm II
				\item [GOA] Grasshopper Optimization Algorithm
				\item [ML] Machine Learning
				\item [CVaR] Conditional Value-at-Risk
				\item [XGBoost] Extreme Gradient Boosting
				\item [MCDM] Multi-Criteria Decision Making
				\item [MINLP] Mixed-Integer Nonlinear Programming
				\item [PPO] Proximal Policy Optimization
				\item[Conv] Converter 
				\item[$x_i(t)$] Output power of energy source $i$ at time $t$ (kW)
				\item[$w_i$] Weighting coefficient for performance metric $i$, where $i=1,2,3,4,\dots$ (dimensionless)	
				\item[$P_{pv}$] Output power of the PV array (kW)
				\item[$Y_{pv}$] Rated power of the PV array under STC (kW)
				\item[$f_{pv}$] PV derating factor
				\item[$\overline{G_t}$] Average solar irradiance on the PV surface (kW/m\textsuperscript{2})
				\item[$G_{t,std}$] Standard irradiance under STC (1 kW/m\textsuperscript{2})
				\item[$\alpha_p$] Temperature coefficient of power (\%/°C)
				\item[$T_c$] PV cell temperature (°C)
				\item[$T_{c,std}$] Cell temperature under STC (25°C)
				\item[$U_H$] Wind speed at the hub height of the wind turbine (m/s)
				\item[$U_A$] Wind speed at the anemometer height (m/s)
				\item[$Z_H$] Hub height of the wind turbine (m)
				\item[$Z_A$] Anemometer height (m)
				\item[$\alpha$] Power law exponent (dimensionless)
				\item[$\rho$] Air density (kg/m\textsuperscript{3})
				\item[$A_S$] Swept area of the wind turbine blades (m\textsuperscript{2})
				\item[$P_{wt}$] Maximum power output of the wind turbine (kW)
				\item[$\alpha_{DG}$] Generator fuel curve intercept coefficient $\left(\frac{\mathrm{L}}{\mathrm{hr} \cdot \mathrm{kW}_{\mathrm{rat}}}\right)$
				\item[$\beta_{DG}$] Generator fuel curve slope coefficient $\left(\frac{\mathrm{L}}{\mathrm{hr} \cdot \mathrm{kW}_{\mathrm{out}}}\right)$
				\item[$P^{rat}_{DG}(t)$] \ Rated DG capacity at time $t$ (kW)
				\item[$P^{out}_{DG}(t)$] \ Generated DG power at time $t$ (kW)
				\item[$Q_1(t)$] Available energy exceeding minimum SOC at time $t$ (kWh)
				\item[$\bar{Q}(t)$] Total stored energy at the initial time step $t$ (kWh)
				\item[$c$] Storage-capacity ratio of the system (dimensionless)
				\item[$k$] Energy storage rate constant (dimensionless)
				\item[$\Delta t$] Time interval under consideration (h)
				\item[$e^{-kt},\ e^{-k\Delta t}$] \  \  \ \ \  \ \  \ \ Exponential terms modeling charge decay over time
				\item[$Q_m(t)$] Total storage capacity at time $t$ (kWh)
				\item[$N_b$] Number of batteries required in the system
				
				\item[$H_{sys}$] Total number of hours the system is designed to operate
				\item[$V_{sys}^{pu}$] Useful voltage of the system per unit
				\item[$Y_{sys}$] System lifespan in years, representing the operational duration of the hybrid system
				\item[$Y_b$] Battery lifespan, indicating the number of years a battery remains functional
				\item[$E_{\text{useable}}^b$] \ Useful energy per battery, defined as the total energy delivered by a single battery over its lifespan				
				\item[$P_{\text{out}}^{conv}(t)$]\ \ \ Output power available after conversion at time $t$ (kW)
				\item[$\eta_{\text{conv}}(t)$] \ Efficiency of the converter at time $t$ (dimensionless)
				\item[$P_{\text{loss}}^{conv}(t)$] \ \ \ \ Power losses during the conversion process at time $t$ (kW)			
				\item[$C_{tot}(t)$] \ Total cost at time $t$ (\$)
				\item[$S(t)$] Salvage value at time $t$, representing the residual value of MG components at end-of-life
				\item[$r$] Discount rate that reflects the time value of money
				\item[$T$] Total project lifetime over which the economic performance is evaluated
				\item[$\lambda$] Scaling factor that modulates the rate of reliability decrease with respect to LPSP
				\item[$P_{load}$] Power demand of the MG at time $t$
				\item[$S(t)$] Power supply of the MG at time $t$
				\item[$P_{useful}(t)$] \ \ \ \ \  Instantaneous useful power delivered to the load at time $t$
				\item[$P_{input}(t)$] \ \ \ Instantaneous total input power from all energy sources at time $t$
				\item[$P_{loss}(t)$] \ \ Instantaneous power losses due to transmission, conversion inefficiencies, and storage losses at time $t$
				\item $P_{g,i}(t)$ is the power supplied by the $i$-th generation source,
				\item $P^{\mathrm{dis}}_{s,j}(t)$ is the discharge power of the $j$-th storage unit,
				\item $N_g$ is the number of generation resources,
				\item $N_s$ is the number of storage systems.
				\item[$E_i$] Annual energy production of component $i$
				\item[$\eta_i$] Efficiency factor of component $i$
				\item[$d_i$] Annual degradation rate of component $i$
				\item[$t$] Year of operation
				\item[$\text{EF}_{\text{grid}}$] Emission factor of grid electricity in kg CO$_2$/kWh
				\item[$E_j$] Annual energy production or consumption of component $j$
				\item[$\text{EF}_j$] Emission factor of component $j$ in kg CO$_2$/kWh				
			\end{description}
		\end{multicols}
	\end{mdframed}
\end{figure*}

\section{Introduction}\label{}
The global energy landscape is undergoing a rapid transformation toward decentralized, sustainable power systems that can meet increasing demands while mitigating environmental impacts \cite{mahuze2025collaborative,zhang2025transactive,bouaouda2024optimal,xu2021bayesian}. Grid-tied community microgrids (MGs) represent a promising solution that enhances reliability and facilitates the integration of renewable resources. They also offer potential economic benefits to local communities \cite{ottenburger2024sustainable,alam2025design,liu2025optimal}. These systems maintain a connection to the main utility grid while providing the capability to operate independently, when necessary. This hybrid approach ensures a balance between resilience and operational efficiency by integrating the grid connectivity with autonomous functionality \cite{ibrahim2024optimal}.
However, designing economically viable community MGs remains challenging due to complex technical requirements, regulatory considerations, and financial constraints \cite{baum2024practical,dinata2024designing,coelho2025monte,valencia2025optimal}. Therefore, cost-effective MG design has attracted much attention in global academic communities.

\subsection{Related work}
In recent years, many studies have focused on cost-effective MG design. In \cite{He2018Techno-economic}, a predominantly renewable energy-based MG was proposed for a residential community in Beijing. The proposed MG demonstrated the capability to supply a minimum of 90\% of the electricity demand utilizing 47-100\% renewable sources. In addition, a moderately sized battery system proved to be more cost-effective than configurations without energy storage.
Liu et al. \cite{Liu2023System} proposed a system parameter design approach for community MGs based on a bi-level optimization model. The proposed system effectively enhances reliability and reduces operation costs without increasing customers' power consumption expenditures. In the proposed approach, the optimal system configuration and operational parameters were generated in an integrated manner.	
A co-optimization strategy was proposed for distributed energy resource planning to minimize the total annualized cost while maximizing fuel savings \cite{Yuan2017Co-Optimization}.

Advanced mathematical formulations have also been applied to optimize distributed energy resource planning.  Lagrange multipliers were used to maximize fuel savings while satisfying the Karush–Kuhn–Tucker  conditions. To minimize the annualized cost, the optimal combination of distributed energy resources was identified using Fourier transform and Particle Swarm Optimization.
Mohamed et al. \cite{Mohamed2019An} proposed an efficient planning algorithm for MGs. The algorithm ensured reliable power flow at minimal cost by determining the optimal grid topology, equipment allocation, and component sizing. It also addressed nonlinear scheduling problems for the installed equipment.
The Harris Hawks Optimization (HHO) algorithm was employed for optimal component sizing of MG \cite{ccetinbacs2021sizing}. The HHO algorithm demonstrated the cost-effectiveness of the design, yielding savings ranging from 1.18\% to 18.23\% of the total NPC. Despite these advantages, environmental impact and efficiency constraints remain unexplored. Considering these aspects, a probabilistic approach was proposed for the optimal sizing and techno-economic assessment of PV-based MGs \cite{Jeyaprabha2023Probabilistic}. However, the scalability of this approach for multi-energy MGs requires further investigation.

To enhance decision-making for critical load management, recent studies have introduced more advanced optimization strategies. In \cite{Alvarez2023Microgrids}, a novel Vectorial MG Optimization method was proposed. This method optimized MG designs for critical loads by balancing the power supply availability, NPC, and power efficiency. As a result, it achieved high reliability and economic viability.
Zhu et al. \cite{Zhu2024Multi-Objective} established a stochastic multi-objective sizing optimization (SMOS) model for MG planning. The proposed model comprehensively incorporated the battery degradation characteristics and the total carbon emissions. A self-adaptive multi-objective genetic algorithm (SAMOGA) was developed to solve the SMOSO model.
In \cite{Oh2024A}, a planning approach was proposed for a gas-electric integrated multi-energy system with an MG. The objective was to enhance power grid resilience against major failures. The method employed mixed-integer linear programming (MILP) to define an operational process based on the required resilience level and to determine the optimal placement of assets within the MG.
To address a multi-energy system, Er et al. \cite{er2024stochastic} employed a two-stage stochastic programming (SP) approach with scenario-based modeling. The proposed design aimed to minimize costs and maximize the system reliability. 

To further advance multi-energy MG planning, recent studies have proposed innovative methods including nature-inspired algorithms and cost-efficient designs. A modified bio-inspired optimization algorithm was employed to design a hybrid energy MG \cite{yan2024effect}. The efficacy of the proposed design was evaluated through a sensitivity analysis.  In \cite{odonkor2025regional}, a cost-effective MG was designed by considering multiple factors, including cost, resilience, and environmental impact. An innovative methodology was proposed to incorporate social dynamics into the design of community MGs \cite{eklund2025evaluating}.  The approach employed MILP to support the decision-making process.
Despite providing valuable insights into cost-effectiveness, most studies have primarily focused on minimizing costs and emissions. Furthermore, the reliability of MG design has not yet been adequately explored. Additionally, there is a paucity of studies in the literature that have considered MG efficiency issues during the design process, which can lead to suboptimal performance of the MG system.

Table \ref{tab:comparison} presents a comparative analysis of recent MG design studies and the proposed framework. The comparison highlights the optimization objectives, treatment of uncertainty, consideration of community MGs, integration of artificial intelligence techniques, and decision-support capabilities. As shown in Table \ref{tab:comparison}, existing studies primarily focus on techno-economic optimization, resilience enhancement, or machine-learning-assisted planning. However, a critical gap persists in addressing economic viability, supply reliability, operational efficiency, and environmental sustainability simultaneously within a community MG context. Furthermore, the integration of preference-based deep reinforcement learning (P-DRL) for adaptive configuration selection remains largely unexplored. This suggests that research on this issue is still in its nascent stages and there are several unresolved challenges that require attention. Therefore, this study proposes a comprehensive design framework for multi-energy community MGs that integrates all the aforementioned factors. %

\begin{table*}[!t]
	\caption{Recent articles related to MG design studies.}
	\label{tab:comparison}
	\centering
	\scriptsize
	\renewcommand{\arraystretch}{1.15}
	\resizebox{\textwidth}{!}{
		\begin{tabular}{|c|c|c|c|c|c|c|c|c|c|c|c|}
			\hline
			\textbf{Ref.} &
			\textbf{Optimization} &
			\textbf{$\mathcal{C} $} &
			\textbf{$ \mathcal{R} $} &
			\textbf{$\eta$} &
			\textbf{$\xi$} &
			\textbf{Uncertainty} &
			\textbf{Sensitivity} &
			\textbf{AI/ML/DRL} &
			\textbf{Community} &
			\textbf{Preference} &
			\textbf{Australian} \\
			\hline
			\cite{eklund2025evaluating} & MILP                     & \checkmark & \checkmark & --         & --         & --         & --         & --  & \checkmark & --      & \checkmark \\
			\cite{zelaschi2025effective} & MILP                & \checkmark & \checkmark & --         & -- & -- & \checkmark & --  & --         & --      & -- \\	
			\cite{fahad2026optimal} & HOMER + MCS          & \checkmark & Partial     & --         & --         & \checkmark & \checkmark & --  & --         & --      & -- \\
			\cite{pradana2026unified} & Unified Robust-SP               & \checkmark & \checkmark & --         & -- & \checkmark & -- & --  & \checkmark        & --      & \checkmark \\
			\cite{tafone2025multi}  & MO-GA + Pareto Search                      & \checkmark & Partial & \checkmark         & \checkmark & -- & -- & -- & \checkmark         & \checkmark      & -- \\
			\cite{oyewole2024multi}  & MILP                     & \checkmark & \checkmark & --         & \checkmark & \checkmark & --         & --  & --         & --      & -- \\
			\cite{ghodusinejad2024multi}  & PESA-II                  & \checkmark & --          & --         & \checkmark & --         & --         & --  & --         & --      & -- \\
			\cite{rehman2025research}  & MILP+MCS                 & \checkmark & \checkmark & --         & \checkmark & \checkmark & --         & --  & --         & --      & -- \\
			\cite{macmillan2024microgrid}  & SP    & \checkmark & \checkmark & --         & --         & \checkmark & --         & --  & --         & --      & -- \\
			\cite{kazemtarghi2024techno}  & ILP                      & \checkmark & \checkmark & --         & --         & Partial    & \checkmark & --  & --         & --      & -- \\
			
			\cite{bukar2025optimal} & GOA+ML                   & \checkmark & \checkmark & --         & --         & Partial    & \checkmark & ML  & --         & --      & -- \\
			\cite{boennec2025robust} & NSGA-II+CVaR             & \checkmark & \checkmark & --         & --         & \checkmark & \checkmark & --  & --         & --      & -- \\
			\cite{bukar2026data} & HOMER+XGBoost+MCDM       & \checkmark & \checkmark & \checkmark & \checkmark & --         & --         & ML  & \checkmark & MCDM    & -- \\			
			\cite{hood2025cost} & MINLP                    & \checkmark & \checkmark & --         & \checkmark & --         & --         & --  & Partial    & --      & -- \\
			\cite{vincent2025sustainable} & SP    & \checkmark & \checkmark & --         & \checkmark & \checkmark & --         & --  & Partial    & --      & -- \\
			\cite{saleheen2026ensuring} & Two-Stage SP       & \checkmark & \checkmark          & -- & -- & \checkmark         & --         & --  & \checkmark         & --      & \checkmark \\
			\cite{alshammari2026energy} & NSGA-II+RL               & \checkmark & \checkmark & --         & \checkmark & \checkmark & \checkmark & RL  & --         & --      & -- \\
			
			\cite{rajendran2026meta} & Meta-RL + PPO               & \checkmark & \checkmark & \checkmark         & \checkmark & -- & -- & DRL  & Partial         & --      & -- \\
			\cite{ezzati2026resilient} & Two-Stage SP              & \checkmark & \checkmark & --         & -- & \checkmark & \checkmark & --  & \checkmark         & --      & -- \\				
			\hline			
			\textbf{This Study} & \textbf{MO-OPT+DRL} &\textbf{\checkmark} & \textbf{\checkmark} & \textbf{\checkmark}	&	\textbf{\checkmark}	 & \textbf{\checkmark} & \textbf{\checkmark} & \textbf{DRL} & \textbf{\checkmark} & \textbf{\checkmark}&\textbf{\checkmark} \\	\hline
		\end{tabular}
	}
\end{table*}

\subsection{Motivation}
\begin{figure}
	\centering
	\includegraphics[width= \linewidth]{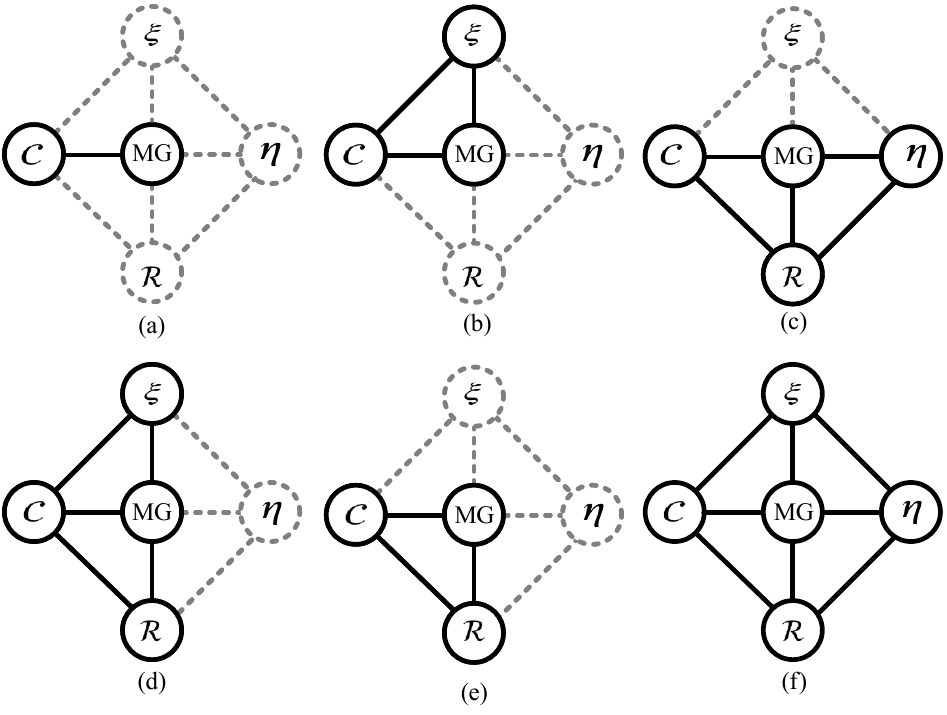}
	\caption []{ Comparative analysis of MG design approaches: (a) MG design prioritizing economic factors alone, (b) MG design integrating both economic  and environmental  considerations, (c) MG Design incorporating economic aspects alongside reliability  and efficiency  metrics, (d) MG design  merging economic considerations with environmental and reliability aspects, (e) MG design emphasizing economic and reliability factors, (f) Proposed MG design framework that synergistically combines economic, environmental, efficiency, and reliability factors for optimized performance.}
	\label{fig:ch4_f1}
\end{figure}

The increasing demand for sustainable, cost-effective, and resilient energy systems has accelerated the  widespread adoption of MGs as viable solutions for decentralized energy management. Despite this progress,  existing MG design methodologies frequently focus on individual performance metrics. These include economic feasibility, environmental sustainability, efficiency, and reliability.  However, they often fail to synergistically integrate all four aspects into a unified framework.
This fragmented approach constrains the potential for MGs to operate optimally in dynamic and uncertain energy environments. A comparative analysis of the existing design approaches is presented in Fig.  \ref{fig:ch4_f1}.

\subsection{Contribution}
This study aims to develop a comprehensive cost-effective design framework for grid-tied community MGs, ensuring economic feasibility and environmental sustainability without compromising reliability and efficiency. The proposed framework has the potential to facilitate financially viable MG deployment for residential communities. In summary, the main contributions and originality of this study are as follows:

\begin{itemize}	 
	\item  A preference-based multi-objective design framework is developed by integrating HOMER-based configuration generation with DRL-assisted solution selection to systematically explore trade-offs among cost, reliability, efficiency, and emissions.	
	\item A rigorous mathematical model is presented to enhance understanding of complex MG systems, incorporating practical operational constraints.
	\item A detailed cost-benefit analysis is performed to evaluate system performance using NPC, LCOE, and reliability metrics.
	\item 	The proposed approach is validated using real-world data and simulations, demonstrating its effectiveness in reducing operational costs while ensuring energy reliability for community-scale applications.	
\end{itemize}

 \section{Problem formulation}
 In this section, a multi-objective optimization (MO-OPT) framework is developed to optimally size the solar photovoltaic (PV) system, wind turbine (WT), diesel generator (DG), grid converter, and battery of a community MG. The formulation considers a constraint on maximum grid capacity. The optimization problem is initially defined, followed by a discussion of the objectives and variables. Subsequently, the developed MO-OPT framework is employed to optimally size the MG components under various operational scenarios, while adhering to the maximum grid capacity limit.

 \subsection{Problem statement}
 The increasing demand for reliable, efficient, and sustainable energy in residential communities necessitates innovative solutions to address the challenges of rising energy costs, environmental impact, and the need for an uninterrupted power supply. Conventional energy systems predominantly rely on fossil fuels, resulting in high operational costs  and substantial carbon emissions ($ \xi$). Furthermore, these systems are susceptible to reliability issues, particularly in remote and underserved areas. Hybrid MGs, which integrate multiple energy sources, offer a promising solution. However, designing such systems to  ensure cost-effectiveness while simultaneously  optimizing reliability, efficiency, and emissions reduction  poses a significant challenge.
  
 \subsection{Decision variables}
 The design of a hybrid microgrid (MG) requires simultaneous decisions regarding both the selection of energy resources and their corresponding installed capacities. Accordingly, the decision space is represented by two complementary variable sets: a binary architecture vector  $\mathbf{y} \in \{0,1\}^n$ and a  capacity decision vector $\mathbf{x} \in \mathbb{R}^n$.
Let $\mathcal{I}$ denote the index set of candidate components, e.g., $\mathcal{I} = \{\text{PV},\, \text{WT},\, \text{DG},\, \text{BESS},\, \text{Conv},\, \text{Grid}\}$. The total number of candidates is $ 	n \triangleq |\mathcal{I}| $. The architectural configuration of the MG is defined through the binary decision vector: 
\begin{equation}
	\mathbf{y}=\{y_i\}_{i\in\mathcal{I}}, \qquad y_i\in\{0,1\},
	\label{eq:y_def}
\end{equation}
 where \(y_i \in \{0, 1\}\). Condition \(y_i = 1\) indicates that the energy source \(i\) is incorporated into the system, whereas \(y_i = 0\) signifies its exclusion. Meanwhile, installed capacity of energy sources is defined by:  
\begin{equation}
	\begin{aligned}
			& \mathbf{x}=\{x_i\}_{i\in\mathcal{I}} , \quad x_i\in\mathbb{R}_{\ge 0}\\
			\text{s.t.} \quad & x_i^{\min} \le x_i \le x_i^{\max},\quad \forall i \in \mathcal{I}.
		\end{aligned}
\end{equation}
To ensure logical consistency between the discrete architecture decision and the continuous sizing decision, $\mathbf{x}$ and $\mathbf{y}$ are coupled using the standard on/off constraints
\begin{equation}
	y_i\,x_i^{\min} \le x_i \le y_i\,x_i^{\max}, \qquad \forall i\in\mathcal{I}.
	\label{eq:coupling}
\end{equation}
Constraint~\eqref{eq:coupling} enforces $x_i=0$ whenever $y_i=0$, while restricting $x_i$ to the feasible sizing interval $[x_i^{\min},\,x_i^{\max}]$ whenever $y_i=1$.

For computational tractability and systematic exploration of the techno-economic design space, each sizing variable is discretized using a predefined step size $\Delta_i>0$. The resulting candidate set for component $i$ is defined as in (4).
\begin{equation}
	\begin{aligned}
		S_i = \Bigl\{ x_i^{\min} + k\Delta_i \;\Bigm|\; 
		& k = 0,1,\ldots, \\
		& \left\lfloor 
		\frac{x_i^{\max} - x_i^{\min}}{\Delta_i}
		\right\rfloor 
		\Bigr\}, 
		\quad \forall i \in \mathcal{I}.
	\end{aligned}
\end{equation}
where $\Delta_i$ denotes the discretization step size. 
The complete configuration domain $\Phi$ is therefore constructed as the Cartesian product of all individual component search spaces. Consequently, $\Phi$ represents the full techno-economic design space explored, where each point corresponds to a unique MG configuration with specific capacity allocations for generation, storage, and conversion units.
\begin{equation}
	\Phi = \prod_{i \in \mathcal{I}} S_i.
	\label{eq:config_domain}
\end{equation}
where each configuration $\theta_j = (\mathbf{x_j}, \mathbf{y_j}) \in \Phi$ denotes the $j$-$^{th}$ candidate MG design characterized by a unique combination of component selection states and installed capacities. The set of feasible MG configurations is defined as
\begin{equation}
	\Phi_f=
	\left\{
	\theta_j\in\Phi
	\;\middle|\;
	\theta_j
	\text{ satisfies all constraints}
	\right\}.
	\label{eq:feasible_set}
\end{equation}
The feasible set $\Phi_f = \{\theta_j\}_{j=1}^{N_f}$ contains all candidate MG configurations that satisfy the power balance requirement, component capacity limits, grid interconnection constraints, reliability requirements, and budget constraints, where $N_f$ denotes the number of feasible configurations.

\subsection{Objective functions formulation}
The proposed sizing tool is developed to determine the optimal capacities of PV, wind, diesel, BESS, and grid components under a composite objective function, $ \mathcal{J}(\theta) $, as defined in Eq. (\ref{eq:vector_cost_function}). Normalization coefficients are employed to ensure that all summation terms have the same order of magnitude.The determination of the MG architecture and sizing of energy resources is contingent upon balancing four critical performance metrics: minimizing the NPC ($ \mathcal{C} $) and carbon emissions ($ \xi  $), while maximizing the power supply reliability ($ \mathcal{R}$) and the system efficiency ($ \eta$). Accordingly, the MG design problem is formulated as a MO-OPT model. Let $\theta \in \Phi_f $ denote a feasible MG configuration generated from the design space defined in Eqs. (1)–(6). The general multi-objective optimization problem is expressed as
\begin{equation}\label{eq:vector_cost_function}
	\min_{\theta \in \Phi_f}\; \mathcal{J}(\theta)
	= \big[ \mathcal{J}_1(\theta), \mathcal{J}_2(\theta), \ldots, \mathcal{J}_K(\theta) \big]^{\mathrm T},\quad \mathcal{J} \in \mathbb{R}^K .
\end{equation}
where $K$ denotes the number of objectives. In this study, $K = 4$, and the objective functions correspond to minimizing the NPC, maximizing system reliability and efficiency, and minimizing carbon dioxide emissions. The individual objective functions are defined as
\[
\begin{aligned}
	\min_{\theta \in \Phi_f}\;&
	\ \mathcal{J}(\theta)=
	\big[  \mathcal{C}(\theta ),\,
	1-\mathcal{R}(\theta ),\,
	1-\eta(\theta ),\,
	\xi(\theta ) \big]^{\mathrm T}
	\in \mathbb{R}^4 .\\
	\text{s.t.} \quad & \sum_{i \in \mathcal{I}} x_i(t)\, y_i \geq P_{\text{load}}(t), \quad \forall t \\
	& 0 \leq x_i \leq y_i \cdot P_{i,\text{max}}, \quad \forall i \in \mathcal{I} \\
	& \sum_i \left(\mathcal{C}_{\text{cap},i} \cdot y_i + \mathcal{C}_{\text{op},i} \cdot x_i + \mathcal{C}_{\text{om},i} \cdot x_i \right) \leq B \\
	& y_i \in \{0,1\}, \quad \forall i \in \mathcal{I}\\
	& x_i \in S_i, \quad \forall i \in \mathcal{I}.
\end{aligned}
\]

To identify the trade-off solutions across all conflicting objectives, each feasible configuration $\theta_j \in \Phi_f$ is evaluated through a performance vector as
\begin{equation}
	\mathcal{J}(\theta_j) = [ \mathcal{C}(\theta_j),\ 1-\mathcal{R}(\theta_j),\ 1-\eta(\theta_j),\ \xi(\theta_j)].
	\label{eq:performance_vector}
\end{equation}

The vector $\mathcal{J}(\theta_j)$ defines the multi-dimensional performance space for all feasible MG designs, forming the basis for multi-objective trade-off analysis.  
To identify trade-off solutions among the conflicting objectives, a configuration $\theta_j$ is considered non-dominated if no other configuration performs better in all objectives simultaneously. Accordingly, the set of non-dominated solutions is defined as
\begin{equation}
\Phi^{*} = \left\{ \theta_j \in \Phi_f \;\middle|\; \nexists \theta_k \in \Phi_f : \mathcal{J}(\theta_k) \prec \mathcal{J}(\theta_j) \right\}.
	\label{eq:pareto_front}
\end{equation}
where $\prec$ denotes Pareto dominance.
The resulting set $\Phi^{*}$ represents the set of non-dominated design alternatives that characterize the trade-off space among cost, reliability, efficiency, and emissions. This Pareto-approximate solution space forms the basis for subsequent preference-based decision analysis.

To ensure comparability among objectives with different physical units, each objective is normalized using min--max normalization:
\begin{equation}\label{eq:normalised-objective}
	\hat{\mathcal{J}}_k = \frac{\mathcal{J}_k - \mathcal{J}_k^{\min}}{\mathcal{J}_k^{\max} - \mathcal{J}_k^{\min}}, 
	\qquad k = 1, \dots, 4
\end{equation}

This normalization procedure is identical to that used in the DRL environment, ensuring consistency between the optimization framework and the learning process. For preference-based decision-making and operational optimization, the normalized objectives are aggregated into a scalar utility function using a weighted-sum formulation:
\begin{equation}\label{EQ:scalar-obj}
	\mathcal{J}_s = \sum_{k=1}^{4} w_k \hat{\mathcal{J}}_k, 
	\qquad \sum_{k=1}^{4} w_k = 1, 
	\qquad w_k \ge 0
\end{equation}

where $w_k$ represents the relative importance of cost, reliability, efficiency, and emissions, respectively. 
The scalarized objective function (\ref{EQ:scalar-obj}) is directly employed as the reward-generation mechanism within the DRL environment. At decision step $t$, the state is represented by the normalized performance vector
 \begin{equation} \label{eq:State vector}
 s_t =  \left[ \hat{\mathcal{C}}_t, \hat{\mathcal{R}}_t, \hat{\eta}_t, \hat{\xi}_t \right]
 \end{equation}
where $\hat{\mathcal{C}}_t$, $\hat{\mathcal{R}}_t$, $\hat{\eta}_t$, and $\hat{\xi}_t$ denote the normalized cost, reliability, efficiency, and emissions metrics associated with the selected MG configuration at decision step $t$, respectively. The reward is then defined as the negative scalarized utility value:
\begin{equation}\label{Reward}
	r_t = - \mathcal{J}_s(s_t)
\end{equation}
where $s_t$denotes the state observed at decision step $t$. The DRL agent learns a policy  $\pi^{*}$ that maximizes the expected cumulative reward:
\begin{equation}\label{Eq:DRL}
	\pi^{*} = \arg\max_{\pi} \, \mathbb{E}_{\pi} \left[ \sum_{t=0}^{T} \gamma^{t} r_t \right]
\end{equation}
where $\gamma \in (0,1]$ is the discount factor and $T$ denotes the episode horizon. After training, the learned policy is employed to evaluate and rank candidate MG configurations according to stakeholder-defined preferences. The final MG configuration is then determined by minimizing the preference-weighted scalarized utility function:
\begin{equation}\label{EQ:Theta}
	\theta^{*} = \arg\min_{\theta_j \in \Phi^*} \sum_{k=1}^{4} w_k \hat{ \mathcal{J}}_k(\theta_j)
\end{equation}
where $\theta^{*}$ denotes the preference-optimal MG configuration, obtained by minimizing the weighted aggregate objective. This solution provides the most desirable balance among net present cost, system reliability, energy conversion efficiency, and lifecycle carbon emissions in accordance with the specified stakeholder preference vector.

\section{Proposed methodology}

The proposed framework adopts a two-stage sequential architecture that integrates physics-based simulation with data-driven decision optimization, enabling systematic multi-objective MG design under realistic operational and stakeholder constraints.

\textbf{Stage 1:} 
In the first stage, HOMER Pro is employed as a high-fidelity techno-economic simulation platform to systematically enumerate and evaluate a comprehensive set of candidate MG configurations. Each configuration is assessed against a multi-dimensional performance criterion encompassing net present cost (NPC), system reliability, energy conversion efficiency, and lifecycle carbon emissions.  An exhaustive grid-search mechanism (Algorithm \ref{Alg_GridSearch}) is first employed to enumerate all feasible system architectures within the prescribed design boundaries. Subsequently, a derivative-free optimization procedure (Algorithm \ref{alg:optimal-configuration}) is executed to identify economically optimal configurations while preserving superior reliability performance. 
The evaluation process is repeated across multiple sensitivity scenarios to capture the effects of uncertainties associated with renewable resource variability, load fluctuations, and economic parameters. Candidate solutions are subjected to rigorous feasibility filtering based on component capacity limits, renewable penetration requirements, reserve margin criteria, network interconnection constraints, and lifecycle cost boundaries.
The output of this stage constitutes a structured Pareto-approximate solution space, wherein each candidate configuration is characterized by a vector of normalized performance metrics serving as the foundational dataset for subsequent optimization.
\begin{algorithm}[tp]
	\caption{Optimized Grid Search for Feasible MG Designs}
	\label{Alg_GridSearch}
	\begin{algorithmic}[1]
		\State \textbf{Input:} Technical, economic, and resource data;
		decision variables defined in Eqs.~(1)--(6).
		item \textbf{Step 1: Define search spaces} 
		Each component $i \in \mathcal{I}$ is assigned a search range 
		$S_i = [\text{min}_i, \text{max}_i, \Delta_i]$, 
		and the overall configuration set is 
		$\Phi= \prod_{i \in \mathcal{I}} S_i.$
		\vspace{0.5em}
		\State \textbf{Step 2: Evaluate Each Configuration.}
		\For{each configuration $\boldsymbol{\theta}_j \in \Theta$}
		\State Simulate annual MG operation		
		\State Perform dispatch and energy-balance calculations		
		\State Check feasibility constraints:		
		\If{$\boldsymbol{\theta}_j$ is feasible }
		\State Compute objective vector $\mathcal{J}(\boldsymbol{\theta}_j)$ using Eq.~(\ref{eq:performance_vector}).
		\State Store $\theta_j$ and $\mathcal J(\theta_j)$
		\EndIf
		\EndFor
		\vspace{0.5em}
		\State \textbf{Step 3: Construct feasible set}
		\[
		\Phi_f=
		\{\theta_j\in\Phi
		\mid
		\theta_j \text{ satisfies all constraints}\}
		\]
		\State \textbf{Output:} Feasible configuration set $\Phi_f$ and
		corresponding performance vectors  $\mathcal{J}(\boldsymbol{\theta}_j)$
	\end{algorithmic}
\end{algorithm}

\textbf{Stage 2:}
The second stage introduces a preference-driven DRL framework (Algorithm \ref{alg:pareto-policy-gradient}) that operates over the solution space generated in Stage 1. Critically, the DRL agent does not exercise direct control over individual MG components or dispatch schedules; rather, it functions as an intelligent decision-support mechanism that learns to evaluate and rank candidate configurations through iterative interaction with a purpose-built optimization environment. The state space encodes the normalized multi-objective performance vectors of candidate configurations, while the action space is discretized to correspond to the selection among feasible configurations identified in Stage 1. Reward signals are derived from a preference-based utility function that explicitly incorporates stakeholder priorities through weighted trade-offs among conflicting objectives. Consequently, the learning process guides the agent toward configurations that best satisfy the desired balance between cost-effectiveness, system reliability, energy efficiency, and emissions reduction.

The sequential coupling of HOMER Pro-based simulation with DRL-based preference evaluation offers several methodological advantages over conventional single-objective or heuristic optimization approaches. By decoupling the physical feasibility assessment from the preference-driven selection process, the framework preserves engineering rigor in system modeling while introducing principled adaptability in the decision layer. This architecture enables the systematic identification of balanced, stakeholder-aligned solutions across a Pareto-approximate front, accommodating the inherently conflicting nature of techno-economic and environmental objectives in real-world MG planning.

\begin{algorithm}[t]
	\caption{Derivative-Free Refinement of MG Configurations}
	\label{alg:optimal-configuration}
	\textbf{Input:}
	Feasible configuration set $\Phi_f$.	
	\begin{algorithmic}[1]		
		\State Select an initial feasible configuration
		$\theta_0\in\Phi_f$		
		\State Evaluate objective vector: $\mathcal J(\theta_0)$		
		\State Set: $\theta_{\mathrm{best}}
		\leftarrow
		\theta_0$			
		\While{stopping criterion is not satisfied}		
		\State Generate candidate configuration
		$\theta_{\mathrm{new}}$		
		\State Simulate MG operation		
		\State Evaluate objective vector $\mathcal J(\theta_{\mathrm{new}})$				
		\If{$\theta_{\mathrm{new}}$ improves the current solution}		
		\State Update:  $\theta_{\mathrm{best}}
		\leftarrow
		\theta_{\mathrm{new}}$		
		\EndIf		
		\EndWhile		
		\State Determine non-dominated solutions according
		to Pareto dominance	 $\Phi^\ast$	using Eq. (\ref{eq:pareto_front}).	
	\end{algorithmic}
	\textbf{Output:}
	Refined non-dominated configuration set $\Phi^\ast$.
\end{algorithm}

\begin{algorithm}[t]
	\caption{Preference-Based DRL Configuration Selection}
	\label{alg:pareto-policy-gradient}
	\begin{algorithmic}[1]		
		\Require Pareto-optimal configuration set $\Phi^*$ generated using Algorithm~2 and corresponding objective vectors $\mathcal{J}(\theta_j)$.		
		\State Normalize objective values according to Eq.~(\ref{eq:normalised-objective}).		
		\State Construct the DRL environment:
		\begin{itemize}
			\item Define state space using the normalized objective vector in Eq.~(\ref{eq:State vector}).
			\item Define the action space as the set of feasible configurations $\theta_j \in \Phi$.
		\end{itemize}		
		\State Compute the preference-based reward according to Eq.~(\ref{Reward}).		
		\State Train the DRL agent to maximize the expected cumulative reward in Eq. (\ref{Eq:DRL})		
		\While{convergence criterion is not satisfied}
		\State Evaluate configurations under the current policy.
		\State Compute cumulative discounted rewards (Eq.~(\ref{Eq:DRL})).	
		\State Update the policy.
		\EndWhile		
		\State Evaluate all feasible configurations $\theta_j \in\Phi^*$.		
		\State Rank configurations according to the learned policy.
		\State Identify the preferred configuration according to Eq.~(\ref{EQ:Theta}).
		\Ensure DRL-selected MG configurations and their corresponding trade-offs among cost, reliability, efficiency, and emissions.		
	\end{algorithmic}
\end{algorithm}

\subsection{Design case study}
\begin{figure}
	\centering
	\includegraphics[width= \linewidth]{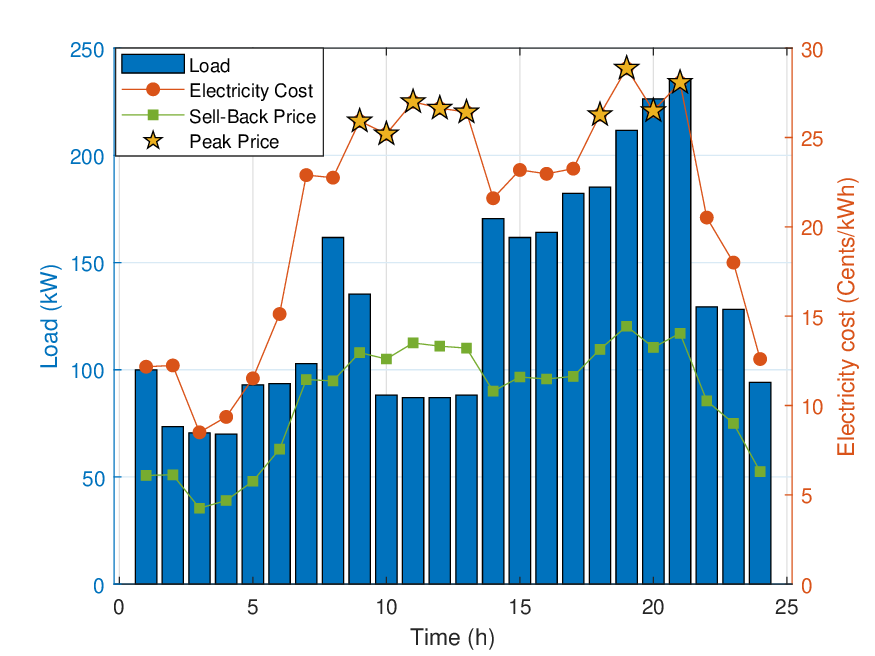}
	\caption []{Hourly electrical demand and electricity price (\$), sellback price for the case srudy community.}
	\label{fig:Load}
\end{figure}

The rural Australian community of Central Tilba in New South Wales, with 288 residents close to the Princes Highway, is selected as the case study for this investigation. This region continues to face challenges in accessing reliable and cost-effective electricity, despite being recognised as one of Australia’s top cultural sites. Power demand analysis of end users and the development of a cost-effective MG model are critical components of this study.The proposed design assumes that the community’s electrical load has not changed substantially in recent years. The average daily power consumption is about 3139.3 kWh/day, and the peak load is 235.2 kW. Fig.  \ref{fig:Load} illustrates the load profile of the rural community. The daily electrical load also follows a typical oscillating distribution, with higher power demand during the day and lower at night, with the peak load concentrating between 5:00 PM to 9:00 PM. Monthly solar radiation profile and average wind speed of Central Tilba, NSW ($36^\circ18.8'S, 150^\circ04.6'E$) are considered for this study. Solar and wind resource data are obtained from the NASA prediction of Worldwide Energy Resources (POWER)  database. The monthly average solar radiation and wind speed for this region are shown in Fig.  \ref{fig:pvwind}.

\subsection{MG element models}
This subsection presents the system component models and the design considerations employed in the proposed approach to calculate the NPC, supply reliability, system efficiency, and carbon reduction of the MG for Central Tilba, NSW, Australia.

\subsubsection{PV module} 
Flat PV panels are selected for this study due to their being readily available on the local market. Detailed information about the PV module is given in Table \ref{tab:tec_eco_data}. The power output of a PV module is estimated considering the temperature influences on it. The following equation is used to calculate the output of the PV module \cite{uddin2023microgrids}:
\begin{equation}\label{Eq_PV} 
	P_{pv}=Y_{pv} f_{pv} \left(\frac{\overline{G}_t}{\overline{G}_{t,std}}\right) \left[1+\alpha_p \left(T_c-T_{c,std} \right) \right].
\end{equation}
\subsubsection{WT model}
Wind speeds at heights greater than 10 m are needed to produce efficient wind energy. However, the data was acquired with a 10 m standard anemometer. Therefore, hub height wind speeds at a potential location are derived using the well-known power law,
\begin{equation} \label{eq:c4_wt1}
	U_H=U_A \times \left(\frac{Z_H}{Z_A} \right)^\alpha
\end{equation}
The exponent value of $ \alpha= $ 0.14, or (1/7), has been widely accepted as an accurate reflection of the prevalent conditions. $U_H$, $\rho$, and $ A_S $, all influence the power production of WT. Therefore, the maximum $ P_{wt} $ can be expressed by the  Eq. (\ref{eq_wind_power}) \cite{shafiullah2012prospects}:
\begin{equation} \label{eq_wind_power}
	P_{wt}=\frac{1}{2} \rho A_S U_{H}^2.
\end{equation}

\subsubsection{DG model}
In this investigation, a 60 kW Caterpillar DG generator is selected. The features of DG are listed in detail in Table \ref{tab:tec_eco_data}. To avoid damaging the generator's exhaust system, the minimum load ratio is set to 25\%. 
The hourly fuel consumption of the DG can be mathematically described using a linear law \cite{uddin2023microgrids}. This model is based on the power required by the load:
\begin{equation}
	\mathcal{C}_{\text{DG}}(t) = \alpha_{\text{DG}} \mathcal{P}^{rat}_{\text{DG}}(t) + \beta_{\text{DG}} \mathcal{P}^{out}_{\text{DG}}(t).
\end{equation}

\subsubsection{BESS model}
In this study, the battery-charging process commences when power generation exceeds consumption. Conversely, when the renewable system is unable to meet energy demands, the storage system supplies the load. The battery state of charge (SOC) is used to establish the charge and discharge limits. Battery capacity is defined as the maximum extractable charge from a fully charged battery. The maximum power output of a storage system can be determined using the following equation \cite{cordero2020optimization}:
\begin{equation}
	\small
	\begin{aligned}
		P_b(t) = \frac{kQ_{\text{m}}(t)e^{-kt} + kQ_1(t)e^{-kt} + \bar{Q}(t)kc(1 - e^{-k\Delta t})}
		{1 - e^{-k\Delta t} + c(k\Delta t - 1 + e^{-k\Delta t})}.
	\end{aligned}
\end{equation}

To ascertain the requisite number of batteries ($ N_{b} $), it is essential to consider factors such as the operational lifespan of the hybrid system and longevity of each individual battery.
\begin{equation}
	N_{\text{b}} = \lceil \frac{H_{sys} \cdot V_{sys}^{pu} \cdot Y_{sys}}{Y_b\cdot E_{\text{usable}}^{\text{b}}} \rceil
\end{equation}

\begin{figure}
	\centering
	\includegraphics[width= \linewidth]{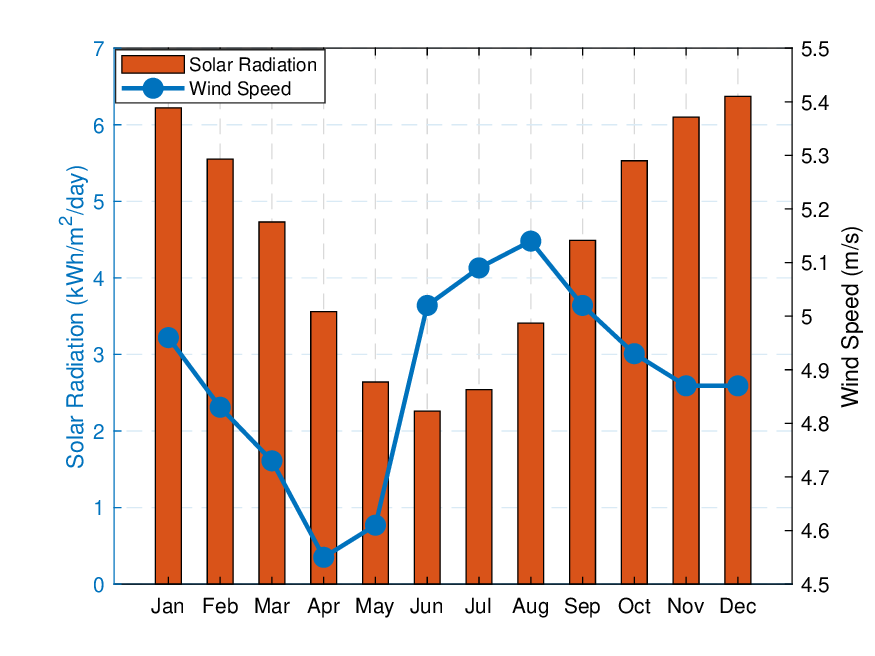}
	\caption []{Daily average solar radiation and wind speed for the case study community.}
	\label{fig:pvwind}
\end{figure}

\subsubsection{Converter model}
The converter model is crucial for the efficient operation of the MG, as it governs the conversion of power between DC and AC forms. The efficiency of this conversion process and the associated power losses are significant factors that impact the overall performance and reliability of the MG system. The output power $ P_{out}^{conv} $ at any given time \( t \) is determined by the following equation:			
\begin{equation}
	P_{out}^{conv}(t) = \eta_{conv}(t) \cdot P_{in}^{conv}(t) - P_{loss}^{conv}(t).
\end{equation}
\begin{table}[t]
	\caption{Technical characteristics and economic data for MG components}
	\label{tab:tec_eco_data}
	\centering
	\renewcommand{\arraystretch}{1.2}
	\resizebox{\columnwidth}{!}{%
	\begin{tabular}{llll}
		\toprule
		\textbf{Component} & \textbf{Characteristic} & \textbf{Value} & \textbf{Unit} \\
		\midrule
		
		\multirow{6}{*}{PV Module} 
		& Nominal power, $P_{\text{pv}}$ & 1 & kW \\
		& Derating factor & 80 & \% \\
		& Capital cost & 1300 & \$/kW \\
		& Replacement cost & 1300 & \$/kW \\
		& O\&M cost & 10 & \$/kW/year \\
		& Lifetime & 20 & Years \\
		
		\midrule
		
		\multirow{7}{*}{Wind Turbine} 
		& Nominal capacity, $P_r$ & 3 & kW \\
		& Cut-in wind speed, $V_{\text{ci}}$ & 4 & m/s \\
		& Cut-out wind speed, $V_{\text{co}}$ & 24 & m/s \\
		& Hub height & 15 & m \\
		& Capital cost & 2300 & \$/kW \\
		& O\&M cost & 207 & \$/kW/year \\
		& Lifetime & 20 & Years \\
		
		\midrule
		
		\multirow{4}{*}{Diesel Generator} 
		& Nominal capacity, $C_{\text{deg}}$ & 60 & kW \\
		& Capital cost & 400 & \$/kW \\
		& Replacement cost & 400 & \$/kW \\
		& O\&M cost & 0.03 & \$/h/kW \\
		
		\midrule
		
		\multirow{8}{*}{Battery Storage} 
		& Nominal capacity, $C_{\text{batt}}$ & 1 & kWh \\
		& Nominal voltage, $V_{\text{batt}}$ & 24 & V \\
		& Roundtrip efficiency, $\eta_{\text{RT}}$ & 90 & \% \\
		& Depth of Discharge (DoD) & 80 & \% \\
		& Capital cost & 700 & \$/kW \\
		& Replacement cost & 700 & \$/kW \\
		& O\&M cost & 10 & \$/year/kWh \\
		& Lifetime & 10 & Years \\
		
		\midrule
		
		\multirow{6}{*}{Power Converter} 
		& Nominal capacity & 1 & kW \\
		& Conversion efficiency, $\eta_{\text{inv}}$ & 95 & \% \\
		& Capital cost & 300 & \$/kW \\
		& Replacement cost & 300 & \$/kW \\
		& O\&M cost & 0 & \$/year \\
		& Lifetime & 15 & Years \\
		
		\bottomrule
	\end{tabular}
	}
\end{table}

\subsection{Performance metrics and evaluation techniques}
Diverse performance metrics and evaluation techniques are used to assess the effectiveness and efficiency of the MG system.  The metrics under consideration include the following aspects. 

\subsubsection{MG economic performance}
Evaluating the economic performance of the MG is a crucial aspect of its overall feasibility and sustainability. The NPC is a widely used metric to assess the total cost of owning and operating a MG over its lifetime. The NPC is calculated using the following equation:		
\begin{equation}
	\mathcal{C} = \sum_{t=0}^{T} \frac{C_{\text{tot}}(t) - S(t)}{(1 + r)^t}.
\end{equation}

The total cost \( \mathcal{C}_{\text{tot}}(t) \)  comprises multiple cost components, each representing a significant financial aspect of the MG operation.		
\begin{itemize}
	\item  \textbf{Capital cost  $ (\mathcal{C}_{cap}) $}:  This represents the initial capital investment required for the establishment of the MG, encompassing equipment procurement, installation, and other associated capital expenditure. 
	\item \textbf{Operation and maintenance cost ($\mathcal{C}_{OM}$}): It encompasses the ongoing operational expenditure essential for the efficient and reliable functioning of the MG, including routine maintenance procedures.
	\item \textbf{Fuel cost  ($ \mathcal{C}_{fuel}(t) $)}:  This accounts for the fuel expenditure required for generators and other fuel-dependent components within the MG system.
	\item \textbf{Replacement cost ($ \mathcal{C}_{rep}(t) $}): This includes the costs associated with replacing or upgrading system components that have reached the end of their operational lifespan over the project duration.
\end{itemize}

By calculating the sum of the discounted total costs and subtracting the salvage values over the lifespan of the project, NPC provides a comprehensive measure of the long-term economic performance of the MG. This metric facilitates the comparison of various design alternatives and enables informed decision-making regarding the viability and cost-effectiveness of MG projects.

\subsubsection{MG power supply reliability}
In this study, the reliability and resilience of MG systems are evaluated by employing the loss of power supply probability (LPSP) indicator as a quantitative metric. The following defines MG reliability as a function of LPSP				
\begin{equation} \label{ch4_rel}
	\mathcal{R} = e^{-\lambda \cdot LPSP}.
\end{equation}
The LPSP quantifies the proportion of the total energy demand that is not met by the MG over a specified period. It measures the probability of power shortages due to insufficient energy generation or storage. LPSP is calculated as follows:
\[ \mathrm{LPSP}
= 100 \,
\frac{\displaystyle \int_{0}^{T} \left[ P_{\mathrm{load}}(t) - S(t) \right]^+ dt}
{\displaystyle \int_{0}^{T} P_{\mathrm{load}}(t) \, dt}
\]

\subsubsection{MG efficiency ($ \eta $)}
The overall MG efficiency ($\eta$) quantifies the capability of the MG to transform the total supplied energy from distributed energy resources, storage systems, and external networks into useful energy consumed by end-users over the scheduling horizon $T$. A higher value of $\eta$ indicates more effective utilization of available energy resources and lower system losses.
\begin{equation}
	\eta
	=
	\frac{
		\int_{0}^{T}
		\left[
		P_{\mathrm{load}}(t)
		-
		P_{\mathrm{unserved}}(t)
		\right]
		\, dt
	}{
		\int_{0}^{T}
		\left[
		\sum_{i=1}^{N_g} P_{g,i}(t)
		+
		\sum_{j=1}^{N_s} P^{\mathrm{dis}}_{s,j}(t)
		+
		P_{\mathrm{imp}}(t)
		\right]
		\, dt
	}
	\times 100\%
\end{equation}

\subsubsection{Environmental impact}	
The environmental benefit of integrating renewable energy sources (PV and wind) into the MG is quantified by the reduction in carbon emissions ($\xi$).  The reduction is calculated using the following equation:
\begin{equation}
	\xi
	=
	\mathrm{EF}_{\mathrm{grid}}
	\sum_{i \in \{\mathrm{PV},\mathrm{WT}\}}
	E_i \, n_i \, (1-d_i)\, t
	-
	\sum_{j \in \{\mathrm{DG},\mathrm{grid}\}}
	E_j \, \mathrm{EF}_j
\end{equation}

	\subsection{Input parameters}
	The cost-effective design of MGs requires a comprehensive set of input parameters that accurately characterize the technical, economic, and operational aspects of the system. These parameters serve as the foundation for optimization algorithms and economic analyses, directly influencing the resulting MG configuration and performance. 
	\begin{itemize}
		\item \textit{Load profiles:} The energy consumption data for the target community is estimated at hourly intervals over one year. This estimation accounts for seasonal variations and peak demand periods. Also, the hourly load with variation of ±10\% is considered throughout the year. 
		\item \textit{Resource availability:} Monthly average values of solar radiation and wind speed for Central Tilba, NSW ($36^\circ18.8'S, 150^\circ04.6'E$), are incorporated into the simulation model. The data, retrieved from the NASA POWER database, serve as key inputs for assessing the site's renewable resource potential. Fig.  \ref{fig:pvwind} shows the corresponding temporal trends.
			
		\item \textit{Grid parameters:} The electricity purchase price from the utility grid is determined under standard operating conditions. The sellback price applies to excess electricity exported to the grid. In addition, the maximum grid capacity for import and export operations influences system performance. Peak pricing during high-demand periods is also incorporated. These parameters substantially influence the economic operation of the MG and the optimal dispatch strategy, particularly in determining the utilization of stored energy versus grid imports.
		\item \textit{Technical and economic parameters:} Table \ref{tab:tec_eco_data} presents the technical specifications and economic data for the principal MG components considered in this investigation. The data represent current market values based on manufacturer specifications and recent industry reports. In addition to the component costs delineated in Table  \ref{tab:tec_eco_data}, this study considers the project financing terms and inflation rates over the project duration (25 years).
	\end{itemize}

	\begin{figure}
	\centering
	\includegraphics[width= \linewidth]{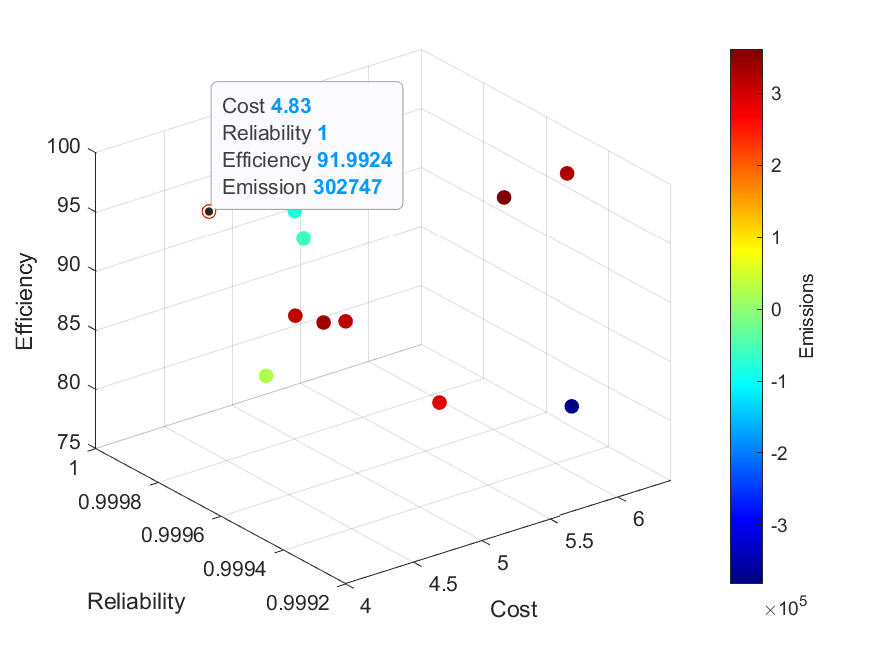}	
	\caption []{Trade-offs among candidate MG configurations selected by the preference-based DRL framework. Each point represents a feasible configuration evaluated using the scalarized multi-objective reward function.}
	\label{fig:ch4_f4}
\end{figure}

\section{Results}
This section provides a detailed evaluation of the simulated MG configurations, revealing the multidimensional trade-offs among economic viability, system reliability, operational efficiency, and environmental impact in grid-connected community MG architectures. 
		\begin{table*}[tp]
			\centering			
			\caption{Comparison of different design approaches}
			\label{tab:design_approach}
			\resizebox{\textwidth}{!}{%
			\begin{tabular}{|l|ccccc|cccc|ccc|}
				\hline
				\multirow{3}{*}{Design Approach (A)} & \multicolumn{5}{c|}{Selected Configuration} & \multirow{3}{*}{ $\mathcal{C}$}  &\multirow{3}{*}{$\mathcal{R} $}    & \multirow{3}{*}{$\eta$ }   & \multirow{3}{*}{$ \xi $} &  \multicolumn{3}{c|}{Other Metrics}  \\ \cline{2-6} \cline{11-13}
				&PV&WT&DG&BES&CON&&&&&LCOE&  $C_{OM}$  &$C_{cap}$ \\ 
				&(kW)&(kW)&(kW)&(kWh)&(kW)&&&&&(\$)&  (\$)  &(M\$) \\ \hline
				Cost-optimized (A1) &451 & 69& 60 &464& 293 &4.47 & 0.9994 & 92.5 & 315909&0.19&189939&1.8 \\
				Reliability-optimized (A2)&785&261&60&872&324 & 5.53 & 1 & 79.744 & 31041& 0.195&183291&2.24\\
				Efficiency-optimized (A3) & 0& 462&60&640&206& 5.81 & 0.9994 & 96.997 & 361807&0.252&243356&1.6 \\
				Emission-optimized (A4) &1057&513&60&1176&549 &6.06 & 0.9993 & 79.956 & -382269&0.175&144142&3.57 \\
				Proposed (A5) & 418&123&0&704&255 &4.83 & 1 & 91.9924& 302747&0.208&201473&1.42 \\
				\hline
			\end{tabular}
	}				
		\end{table*}

		\subsection{Optimal solutions}
		The MO-OPT framework developed in this study provides valuable insights into the trade-offs among cost, reliability, efficiency, and emissions in design of a cost-effective grid-tied community MG. To investigate the trade-offs among competing objectives, five representative solutions were selected from the multi-objective trade-off space. Each solution corresponds to a distinct objective preference scenario. The candidate MG configurations generated by HOMER were subsequently evaluated using the proposed preference-based DRL framework. The DRL agent learns to rank and select configurations according to the scalarized reward function. The resulting solutions reflect different trade-offs among economic, reliability, efficiency, and environmental objectives.	The optimization results shown in Fig.  \ref{fig:ch4_f4} and Table \ref{tab:design_approach}, demonstrate the impact of various design strategies on system performance and economic viability.
		The proposed design (A5) achieves a NPC of \$4.83M, 100\% reliability, and 91.99\% efficiency while maintaining a moderate annual carbon emissions ($ \xi $) level of 302,747 kg/year. These results indicate that the selected MG configuration provides an optimal balance across all considered objectives. The final system configuration comprises 418 kW of PV  capacity, 123 kW of wind power, and 704 kWh of battery storage, completely eliminating reliance on DGs. The design achieves a LCOE  of \$0.208/kWh. 
		
		\begin{figure}
			\centering
			\includegraphics[width= \linewidth]{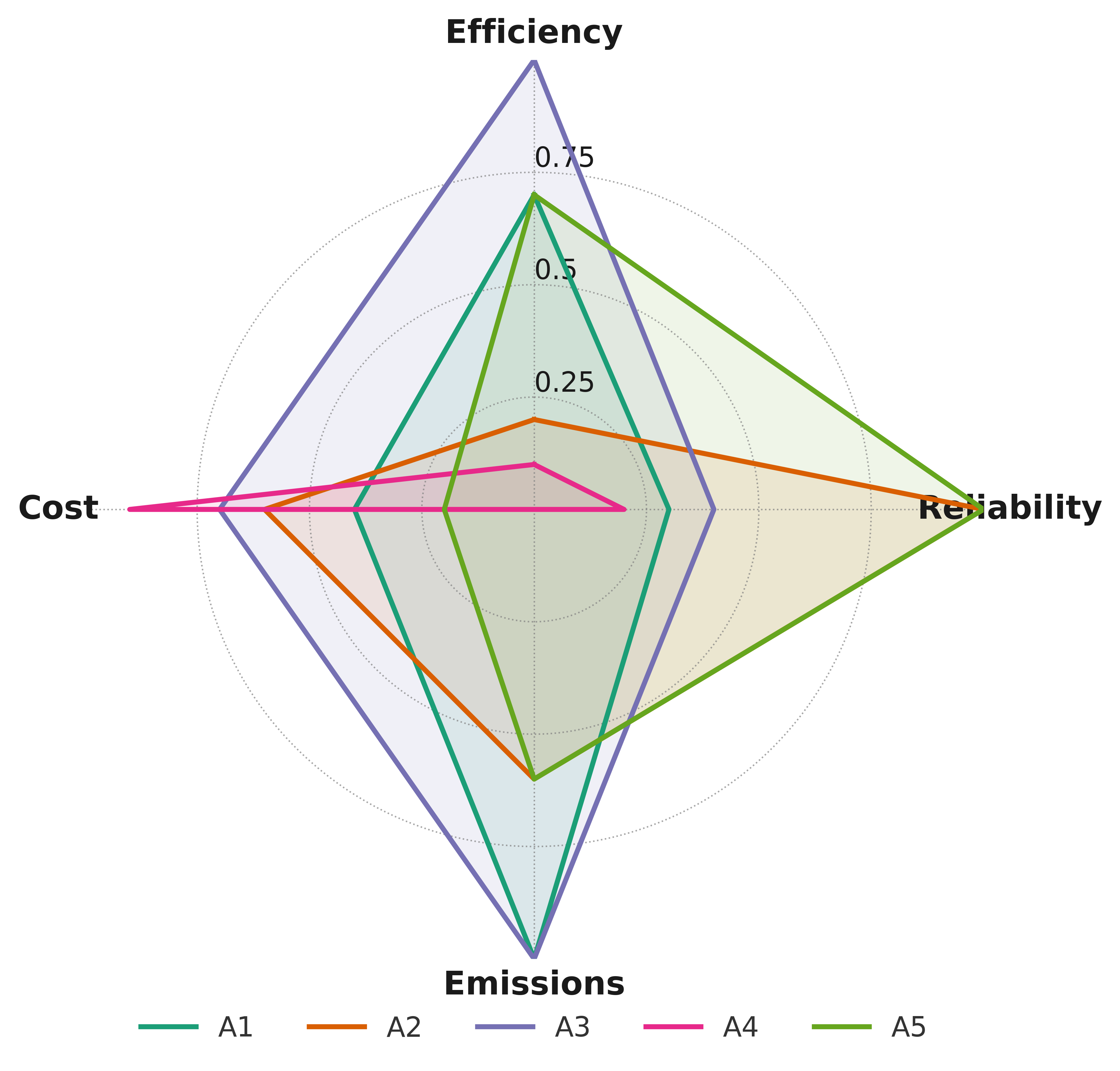}
			\caption []{Normalized multi-objective performance of the five representative optimal MG configurations (A1–A5). The results illustrate the inherent trade-offs among economic, reliability, efficiency, and environmental objectives, highlighting the balanced performance achieved by the proposed design (A5).} 
			\label{fig:f6}
		\end{figure}
		
		\subsection{Comparison with conventional designs}
		A comparative analysis with alternative design configurations provides valuable insights into cost-effective MG architectures. As depicted in Fig.  \ref{fig:f6}, various design strategies prioritize distinct objectives, resulting in diverse system configurations and performance outcomes across cost, reliability, efficiency, and emissions. The findings are discussed in the subsequent subsections.

		\subsubsection{Cost-reliability trade-off} The cost-optimized design (A1) achieves the lowest NPC (\$4.47M) but compromises reliability ($ \mathcal{R} = 0.9994 $). In contrast, the reliability-optimized design (A2) ensures complete reliability at a substantially higher cost (\$5.53M), driven by the increased storage capacity requirement (872 kWh). These findings demonstrate that attaining 100\% reliability leads to an exponential increase in costs, primarily attributed to the necessity for additional backup capacity.
		\subsubsection{Efficiency vs. cost relationship}  The efficiency-optimized design (A3) achieves the highest efficiency (96.997\%), resulting in a higher NPC of \$5.81M. This configuration relies predominantly on wind generation (462 kW) and battery storage (640 kWh). This exemplifies the trade-off wherein maximizing efficiency requires increased investments in storage and renewable energy capacity, which may not always be economically viable.		\subsubsection{Emissions reduction and economic viability} The emission-optimized design (A4) achieves the lowest emissions (-382,269 kg), albeit at the highest NPC (\$6.06M) and the lowest efficiency (79.96\%). This configuration heavily depends on renewable energy integration, comprising 1,057 kW PV and 513 kW wind. It achieves  significant carbon emission reductions but requires a substantial capital investment of \$3.57M. In contrast, the proposed design (A5) achieves a 4.83\% cost reduction while maintaining acceptable emissions and high efficiency, rendering it more feasible for practical applications.
		\subsubsection{Proposed design} The proposed configuration (A5) demonstrates an optimal balance across all objectives, exhibiting 100\% reliability, 91.99\% efficiency, and moderate emissions of 302,747 kg/year. Compared to alternative approaches, it presents a well-balanced integration of renewable energy sources while maintaining economic viability and environmental benefits. The LCOE  for the proposed design (\$0.208/kWh) is lower than that of the efficiency and emission-optimized designs, ensuring economic feasibility. Furthermore, its capital expenditure of \$1.42M is lower than most other configurations, underscoring its cost-effectiveness in comparison to designs with high renewable energy penetration.

\begin{table}[t]
	\centering
	\caption{Performance variation of the selected optimal MG design under different uncertainties}
	\label{tab:performance_variation}
	\renewcommand{\arraystretch}{1.2}
	\resizebox{\columnwidth}{!}{%
		\begin{tabular}{cccccc}				
			\hline
			\setlength{\arrayrulewidth}{0.4pt}
			\multirow{2}{*}{\textbf{Parameter}} & \multirow{2}{*}{\textbf{Uncertainty [\%]}} & \multicolumn{4}{c}{\textbf{Metrics Deviation [\%]}} \\
			\cline{3-6}
			& & $\mathcal{C}$ & $\mathcal{R}$ & $\eta$ & $\xi$ \\
			\hline
			\multirow{4}{*}{Load}
			& $-5$  & -5.15 & 0    & -0.81 & -15.19 \\
			& $5$   & 2.12 & 0    & 0.07  & 6.95 \\
			& $-10$ & -8.66 & 0    & -1.31 & -26.20 \\
			& $10$  & 5.87  & 0    & 0.45  & 18.06 \\
			\hline
			\multirow{4}{*}{PV Output}
			& $-5$  & -1.02 & 0 & 0.59  & -0.05 \\
			& $5$   & -2.44 & 0 & -1.40 & -7.93 \\
			& $-10$ & 0.00  & 0 & 1.56  & 4.50 \\
			& $10$  & -3.05 & 0 & 2.44  & -11.30 \\
			\hline
			\multirow{4}{*}{Wind Output}
			& $-5$  & -0.40 & 0 & -0.31 & 0.34 \\
			& $5$   & -3.05 & 0 & -0.46 & -8.81 \\
			& $-10$ & 0.61  & 0 & -0.24 & 4.56 \\
			& $10$  & -4.20 & 0 & -0.53 & -13.61 \\
			\hline 
		\end{tabular}			
	}		
\end{table}

		\subsection{Sensitivity to key parameters}
		
		The performance of the proposed cost-effective MG design is evaluated under various system conditions to assess its reliability, efficiency, economic feasibility, and environmental impact. This section presents the key findings, including the system response to uncertainties in load demand and renewable energy generation, as well as the sensitivity of the LCOE  to critical economic parameters. The results elucidate the trade-offs involved in optimizing cost, reliability, and sustainability, while ensuring the resilience of the MG against operational fluctuations.
		\subsubsection{Sensitivity analysis of system performance under uncertainty}
\begin{figure*}
	\centering
	\includegraphics[width= \linewidth]{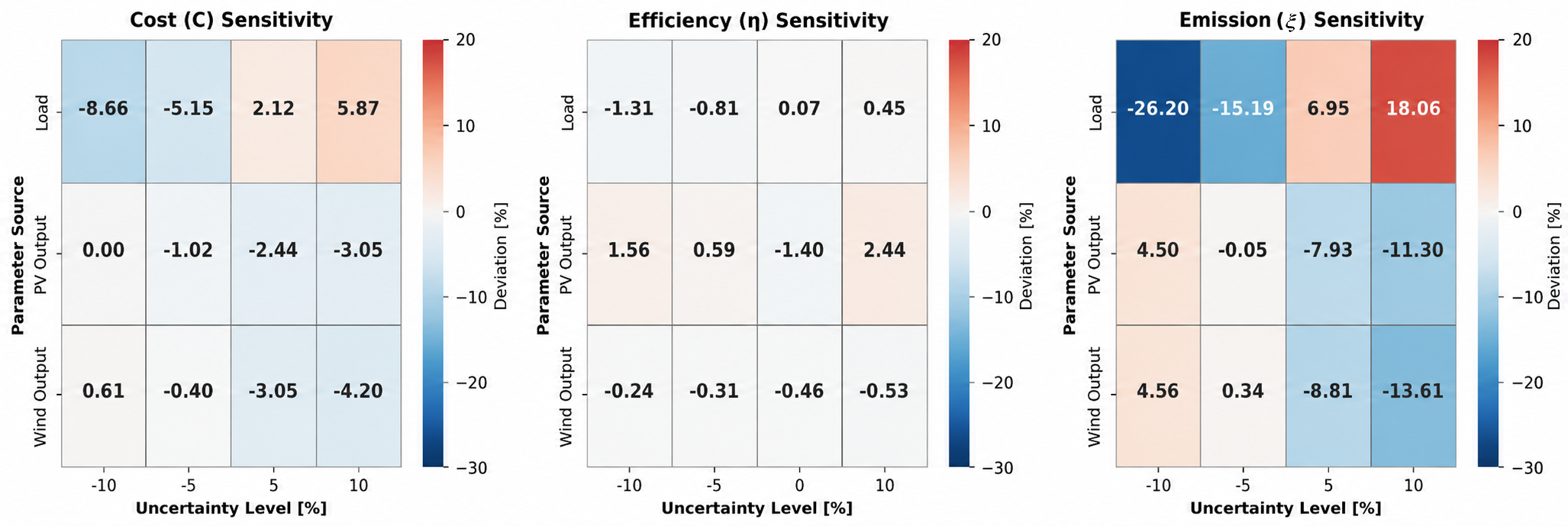}
	\caption []{Detailed sensitivity analysis of the proposed MG under uncertainty. The heatmaps illustrate percentage deviations in (a) NPC, (b) system efficiency, and (c) carbon emissions resulting from variations in load demand, photovoltaic generation, and wind power output over uncertainty levels ranging from -10\% to +10\%.}
	\label{fig:heatmap}
\end{figure*}

		The effects of uncertainty in the load demand, PV output, and wind output on the key MG performance	indicators are summarized in Table \ref{tab:performance_variation} and Fig.  \ref{fig:heatmap}. The evaluated metrics include NPC ($\mathcal{C}$), reliability ($\mathcal{R} $), efficiency ($ \eta $), and carbon emissions ($ \xi $). Table  \ref{tab:performance_variation} quantifies the percentage deviations of these performance indicators under varying uncertainty scenarios, whereas Fig.  \ref{fig:heatmap} provides a comparative visualization of the relative sensitivity of each metric to the considered uncertainty sources. Collectively, the results demonstrate the robustness of the proposed MG design, highlighting its capability to maintain favorable economic, technical, and environmental performance despite fluctuations in renewable resource availability and demand conditions.

		\begin{itemize}
			\item \textbf{Load variations:} The system demonstrates a substantial cost and emission impact due to load fluctuations. A 10\% increase in load results in a 5.87\% increase in NPC and an 18.06\% increase in emissions, as additional energy is required from storage or grid purchases. Conversely, a 10\% reduction in load leads to an 8.66\% reduction in NPC and a 26.2\% decrease in emissions, illustrating the direct correlation between demand and operational sustainability. The reliability remains unaffected ($\mathcal{R} $= 1), indicating that the system maintains full reliability under demand variations. However, efficiency marginally decreases under lower loads due to the underutilization of generation assets.
			\item \textbf{PV output variations:} A 10\% increase in PV output results in a 3.05\% reduction in NPC and an 11.3\% reduction in emissions, demonstrating the economic and environmental advantages of increased solar generation. Conversely, a 10\% reduction in PV output leads to a 3.05\% increase in NPC and an 11.3\% increase in emissions, indicating a greater dependence on backup generation. The efficiency improves with increased PV generation, while the reliability remains constant, further substantiating the robustness of the system.
			\item \textbf{Wind output variations:} The impact of wind fluctuations is less pronounced than that of PV variations, with a 10\% increase in the NPC by 4.2\% and emissions by 13.61\%. A 10\% decrease in wind output results in a modest 0.61\% increase in NPC, suggesting that wind availability has a relatively moderate impact on the overall cost compared to PV fluctuations. The efficiency exhibits a slight decline with lower wind penetration, while the reliability remains consistent.
		\end{itemize}
		These findings indicate that the proposed MG demonstrates resilience to fluctuations in renewable energy generation. However, the optimization of PV penetration is more crucial than wind in ensuring cost-effectiveness and emission reduction. Demand-side management strategies can further mitigate the impact of load variations on the system sustainability.

		\subsubsection{Sensitivity analysis of LCOE}
\begin{figure}
			\centering
			\includegraphics[width= \linewidth]{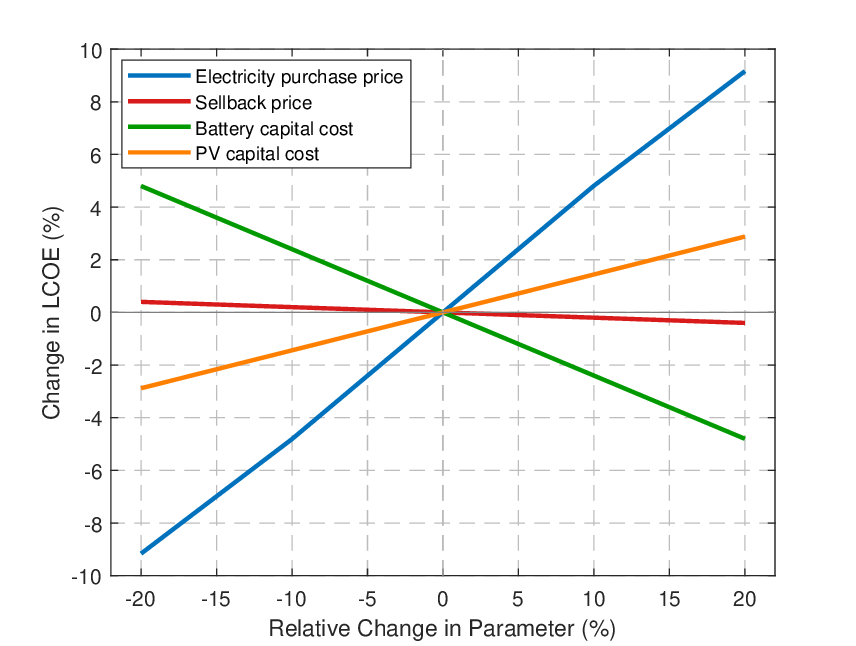}
			\caption []{Sensitivity of the proposed MG's LCOE to key economic parameters. The analysis highlights the relative influence of electricity purchase price, sellback price, battery capital cost, and PV capital cost on system economics and cost competitiveness.}
			\label{fig:ch4_f6}
		\end{figure}

		the sensitivity of the LCOE  to key economic parameters is illustrated in Fig.  \ref{fig:ch4_f6}. The economic parameters considered in this test include electricity purchase price, sellback price, battery capital cost, and PV  capital cost. 	This analysis reveals the relative influence of each parameter on system economics and identifies the components most critical to cost optimization.			
		\begin{itemize}
			\item \textbf{Electricity purchase price:} The LCOE exhibits significant sensitivity to the electricity purchase price, with a 20\% increase resulting in an approximately 10\% increase in the LCOE. This observation suggests that grid electricity costs remain a predominant factor in the overall MG economics, reinforcing the need for self-sufficiency through optimized renewable energy integration.
			\item \textbf{Sellback price:} The sellback price demonstrates a relatively minor influence on LCOE, implying that revenue from excess energy exports contributes little to overall cost-effectiveness.
			This observation emphasizes the necessity for efficient on-site utilization and storage mechanisms rather than dependence on feed-in tariffs.
			\item \textbf{Battery and PV capital costs:} Battery capital costs demonstrate a more pronounced influence on LCOE compared to PV capital costs, as evidenced by the steeper slope in Fig.  \ref{fig:ch4_f6}. A 20\% increase in battery costs results in an approximately 5\% increase in the LCOE, whereas PV cost variations yield a more gradual LCOE change. This observation suggests that battery storage optimization is critical for ensuring cost-effective MG operation, particularly in scenarios with high renewable energy penetration.
		\end{itemize}

		\section{Discussion}
		The results of this investigation demonstrate that an economically viable MG design can be achieved through a strategically balanced optimization approach, ensuring high reliability, efficiency, and emission control without disproportionate cost increases. The proposed approach provides a pragmatic and scalable design methodology for sustainable and economically viable MG operations, and serves as a benchmark for policymakers and energy planners. Although the optimization findings reveal significant technical and financial benefits, several practical aspects of the implementation require further attention.
		
		The optimization results reveal that the proposed configuration achieves complete reliability without diesel generation. This outcome raises important practical considerations regarding the role of battery storage, system resilience during prolonged outages, and long-term economic viability.
		\subsection{Exclusion of DGs and the role of large BESS}
		A crucial strategic decision in this study involves opting for a substantial BESS instead of DGs to guarantee complete reliability in the suggested MG setup. Although BESS provide substantial environmental and operational benefits by eliminating fossil fuel dependency, the findings indicate that high BESS capacity introduces trade-offs in terms of cost, flexibility, and long-term operational resilience during grid outages.
		
		\subsubsection{Operational response time} 
		DGs provide rapid response capabilities, particularly in high-load demand scenarios, where immediate power availability is critical. DGs can achieve full power output within seconds, rendering them suitable for managing sudden load surges and transient events. In contrast, battery storage systems rely on inverter-based power electronics, which may introduce minor delays in the response time, particularly under high-power ramping conditions.
		
		\subsubsection{Long-term grid outage support}
		During prolonged grid outages, the BESS is constrained by finite energy storage capacity. To maintain essential loads, the BESS requires adequate renewable energy generation and frequent recharge cycles. Although BESS can efficiently handle short-duration outages, however, their capacity limitations make them less viable for extended backup operations. Conversely, DGs can provide continuous power for prolonged periods, contingent on fuel availability, making them a more suitable option in regions with unreliable renewable resource availability.
		
		Despite these advantages, in this study, the diesel-free approach is selected to mitigate emissions, reduce operational and maintenance costs, and enhance renewable energy penetration.

		\subsection{Economic impact of large BESS capacity} 
		Achieving 100\% reliability without relying on DG support requires a substantially larger BESS capacity, as observed in the proposed system (704 kWh). However, increased BESS sizing leads to higher NPC and LCOE, as evidenced by the sensitivity analysis (Table \ref{tab:design_approach}). The results indicate that an increase in BESS size directly correlates with higher capital expenditure, consequently increasing the LCOE and overall system cost, thus necessitating an optimal balance between storage sizing and economic viability. Furthermore, oversizing of the BESS may result in underutilized storage capacity during normal grid operation, thereby reducing the cost-effectiveness of the system. 
		The proposed DRL framework is specifically designed for preference-based evaluation and selection of candidate MG configurations rather than direct real-time operational control. Specifically, the DRL agent learns stakeholder preferences to evaluate, rank, and identify the most suitable configuration from the feasible solution space generated by HOMER Pro. Future work will extend this framework to real-time energy management, where DRL agents directly optimize dispatch decisions for battery storage, renewable generation, and grid power exchange under uncertain operating conditions.
		
		\subsection{Implications for future MG design}
		A potential hybrid approach is integrating a small DG unit with a BESS. This configuration could provide a cost-effective and reliable alternative by optimizing both short-term responsiveness and long-term backup capability, while also mitigating emissions.  Future research should explore optimal DG-BESS hybrid configurations to achieve a balance between reliability, cost, and environmental performance. In addition, hydrogen-based backup solutions may offer long-duration energy storage with zero emissions as alternatives to conventional DGs.

\section{Conclusion}

 This study presents a cost-effective MG design that optimally balances the economic feasibility, reliability, efficiency, and environmental impact through a MO-OPT approach. Findings reported in this study shed new light on  developing a comprehensive framework to supply secure, reliable, and affordable power in regional and remote areas.   
The results reveal that the proposed design achieves a $ \mathcal{C}$ of \$4.83M, ensuring high reliability reliability ($ \mathcal{R}$=1).  The results also demonstrate that $ \eta$ of the proposed MG configuration is 91.99\%, while maintaining carbon emissions at 302,747 kg/year. The economic evaluation further demonstrates that the proposed system attains a LCOE  of \$0.208/kWh, with an annual operational expenditure of \$201,473 and capital investment of \$1.42M, rendering it more economically viable than alternative design approaches.
The sensitivity analysis shows that load variations exert the most substantial influence on system emissions and cost, whereas PV  output fluctuations impact economic and environmental performance to a greater extent than wind variations. Furthermore, the LCOE  sensitivity analysis indicates that the electricity purchase price and battery capital costs constitute the most critical factors affecting the overall system cost, reinforcing the necessity for efficient energy management strategies and optimized storage integration.

Therefore, incorporating energy management as a future extension of this study would be valuable, as it could further enhance MG resilience under real-time operational uncertainties.


\section*{Acknowledgment}

The authors 
acknowledge University of New South Wales (UNSW) for providing the financial supports to perform this research.










\printcredits
%

\bibliographystyle{model1-num-names}
%
\bibliography{ref}

@article{shafiullah2012prospects,
	title={Prospects of renewable energy--a feasibility study in the {Australian} context},
	author={Shafiullah, GM and Amanullah, MTO and Ali, ABM Shawkat and Jarvis, Dennis and Wolfs, Peter},
	journal={Renewable Energy},
	volume={39},
	number={1},
	pages={183--197},
	year={2012},
	publisher={Elsevier}
}

@article{uddin2023microgrids,
	title={Microgrids: A review, outstanding issues and future trends},
	author={Uddin, Moslem and Mo, Huadong and Dong, Daoyi and Elsawah, Sondoss and Zhu, Jianguo and Guerrero, Josep M},
	journal={Energy Strategy Reviews},
	volume={49},
	pages={101127},
	year={2023},
	publisher={Elsevier}
}

@article{cordero2020optimization,
	title={Optimization of an off-grid hybrid system using lithium ion batteries},
	author={Cordero, Paul Ar{\'e}valo and Garc{\'\i}a, Juan Lata and Jurado, Francisco},
	journal={Acta Polytechnica Hungarica},
	volume={17},
	number={3},
	pages={185--206},
	year={2020}
}

@article{He2018Techno-economic,
	title={Techno-economic potential of a renewable energy-based microgrid system for a sustainable large-scale residential community in {B}eijing, {C}hina},
	author={He, Li and Zhang, Shiyue and Chen, Yizhong and Ren, Lixia and Li, Jing},
	journal={Renewable and Sustainable Energy Reviews},
	volume={93},
	pages={631--641},
	year={2018},
	publisher={Elsevier}
}

@inproceedings{Liu2023System,
	title={System Parameter Design for Community Microgrid Energy System Based on a Bi-Level Optimization Model},
	author={Liu, Jiangshan and Bi, Youyi},
	booktitle={ASME International Mechanical Engineering Congress and Exposition},
	volume={87646},
	pages={V007T08A052},
	year={2023},
	organization={American Society of Mechanical Engineers}
}

@article{Yuan2017Co-Optimization,
	title={Co-Optimization Scheme for Distributed Energy Resource Planning in Community Microgrids},
	author={C. Yuan and M. Illindala and A. Khalsa},
	journal={IEEE Transactions on Sustainable Energy},
	year={2017},
	volume={8},
	pages={1351-1360},
	doi={10.1109/TSTE.2017.2681111}
	}

@article{Mohamed2019An,
	title={An Efficient Planning Algorithm for Hybrid Remote Microgrids},
	author={Sayed Mohamed and M. Shaaban and Muhammad Ismail and E. Serpedin and K. Qaraqe},
	journal={IEEE Transactions on Sustainable Energy},
	year={2019},
	volume={10},
	pages={257-267},
	doi={10.1109/TSTE.2018.2832443}
	}

@article{ccetinbacs2021sizing,
	title={Sizing optimization and design of an autonomous AC microgrid for commercial loads using Harris Hawks Optimization algorithm},
	author={{\c{C}}etinba{\c{s}}, {\.I}pek and Tamy{\"u}rek, B{\"u}nyamin and Demirta{\c{s}}, Mehmet},
	journal={Energy Conversion and Management},
	volume={245},
	pages={114562},
	year={2021},
	publisher={Elsevier}
}

@article{Jeyaprabha2023Probabilistic,
	title={Probabilistic Techno-Economic Design of Isolated Microgrid},
	author={S. Jeyaprabha and J. Milanović},
	journal={IEEE Transactions on Power Systems},
	year={2023},
	volume={38},
	pages={5267-5277},
	doi={10.1109/TPWRS.2022.3216386}
	}

@article{Alvarez2023Microgrids,
	title={Microgrids Multiobjective Design Optimization for Critical Loads},
	author={Jorge Alejandro May Alvarez and Ignacio Galiano Zurbriggen and Francisco Paz and M. Ordonez},
	journal={IEEE Transactions on Smart Grid},
	year={2023},
	volume={14},
	pages={17-28},
	doi={10.1109/TSG.2022.3195989}
	}

@article{Zhu2024Multi-Objective,
	title={Multi-Objective Sizing Optimization Method of Microgrid Considering Cost and Carbon Emissions},
	author={Xiang Zhu and Guangchun Ruan and Hua Geng and Honghai Liu and Mingfei Bai and Chao Peng},
	journal={IEEE Transactions on Industry Applications},
	year={2024},
	volume={60},
	pages={5565-5576},
	doi={10.1109/TIA.2024.3395570}
	}

@article{Oh2024A,
	title={A Bi-Level Approach for Networked Microgrid Planning Considering Multiple Contingencies and Resilience},
	author={Byeong-Chan Oh and Yeong-Geon Son and Dongbo Zhao and Chanan Singh and Sung-Yul Kim},
	journal={IEEE Transactions on Power Systems},
	year={2024},
	volume={39},
	pages={5620-5630},
	doi={10.1109/tpwrs.2023.3344661}
	}

@article{baum2024practical,
	title={Practical considerations for the design and control of networked microgrids: Enabling effective operation},
	author={Baum, Jackie and Curtiss, Peter and Lee, James and Higginson, Michael and Harwig, Bob},
	journal={IEEE Electrification Magazine},
	volume={12},
	number={2},
	pages={22--32},
	year={2024},
	publisher={IEEE}
}

@article{er2024stochastic,
	title={Stochastic optimal design of a rural microgrid with hybrid storage system including hydrogen and electric cars using vehicle-to-grid technology},
	author={Er, Gulfem and Soykan, Gurkan and Canakoglu, Ethem},
	journal={Journal of Energy Storage},
	volume={75},
	pages={109747},
	year={2024},
	publisher={Elsevier}
}

@article{yan2024effect,
	title={Effect of various design configurations and operating conditions for optimization of a wind/solar/hydrogen/fuel cell hybrid microgrid system by a bio-inspired algorithm},
	author={Yan, Caozheng and Zou, Yunhe and Wu, Zhixin and Maleki, Akbar},
	journal={International Journal of Hydrogen Energy},
	volume={60},
	pages={378--391},
	year={2024},
	publisher={Elsevier}
}

@article{ottenburger2024sustainable,
	title={Sustainable urban transformations based on integrated microgrid designs},
	author={Ottenburger, Sadeeb S and Cox, Rob and Chowdhury, Badrul H and Trybushnyi, Dmytro and Omar, Ehmedi Al and Kaloti, Sujay A and Ufer, Ulrich and Poganietz, Witold-R and Liu, Weijia and Deines, Evgenia and others},
	journal={Nature Sustainability},
	volume={7},
	number={8},
	pages={1067--1079},
	year={2024},
	publisher={Nature Publishing Group UK London}
}

@article{odonkor2025regional,
	title={Regional Performance Analysis of Residential Microgrids: A Multi-Factor Assessment of Cost, Resilience, and Environmental Impact},
	author={Odonkor, Philip and Ashmore, Samuel},
	journal={Energy and Buildings},
	pages={115433},
	year={2025},
	publisher={Elsevier}
}

@article{dinata2024designing,
	title={Designing an optimal microgrid control system using deep reinforcement learning: A systematic review},
	author={Dinata, Noer Fadzri Perdana and Ramli, Makbul Anwari Muhammad and Jambak, Muhammad Irfan and Sidik, Muhammad Abu Bakar and Alqahtani, Mohammed M},
	journal={Engineering Science and Technology, an International Journal},
	volume={51},
	pages={101651},
	year={2024},
	publisher={Elsevier}
}

@article{bouaouda2024optimal,
	title={An optimal sizing framework of a microgrid system with hydrogen storage considering component availability and system scalability by a novel approach based on quantum theory},
	author={Bouaouda, Anas and Sayouti, Yassine},
	journal={Journal of Energy Storage},
	volume={92},
	pages={111894},
	year={2024},
	publisher={Elsevier}
}

@article{coelho2025monte,
	title={Monte Carlo simulation of community microgrid operation: Business prospects in the {B}razilian regulatory framework},
	author={Coelho, Francisco CR and Assis, Fernando A and Jos{\'e} Filho, C Castro and Donadon, Antonio R and Roncolatto, Ronaldo A and Andrade, Vittoria EMS and Rosas, Pedro AC and Barcelos, Silvangela LSL and Saavedra, Osvaldo R and Bento, Rafael G and others},
	journal={Utilities Policy},
	volume={92},
	pages={101856},
	year={2025},
	publisher={Elsevier}
}

@article{ibrahim2024optimal,
	title={Optimal multi-objective sizing of renewable energy sources and battery energy storage systems for formation of a multi-microgrid system considering diverse load patterns},
	author={Ibrahim, Nurul Nadia and Jamian, Jasrul Jamani and Rasid, Madihah Md},
	journal={Energy},
	volume={304},
	pages={131921},
	year={2024},
	publisher={Elsevier}
}

@article{alam2025design,
	title={Design and operation of future low-voltage community microgrids: An {AI}-based approach with real case study},
	author={Alam, Md Morshed and Hossain, MJ and Zamee, Muhammad Ahsan and Al-Durra, Ahmed},
	journal={Applied Energy},
	volume={377},
	pages={124523},
	year={2025},
	publisher={Elsevier}
}

@article{zhang2025transactive,
	title={A transactive energy cooperation scheduling for hydrogen-based community microgrid with refueling preferences of hydrogen vehicles},
	author={Zhang, Xiao-Yan and Wang, Cenfeng and Xiao, Jiang-Wen and Wang, Yan-Wu},
	journal={Applied Energy},
	volume={377},
	pages={124582},
	year={2025},
	publisher={Elsevier}
}

@article{mahuze2025collaborative,
	title={Collaborative optimization framework for capacity planning of a prosumer-based peer-to-peer electricity trading community},
	author={Mahuze, Richard A and Amadeh, Ali and Yuan, Bo and Zhang, K Max},
	journal={Applied Energy},
	volume={384},
	pages={125289},
	year={2025},
	publisher={Elsevier}
}

@article{valencia2025optimal,
	title={Optimal planning and management of the energy--water--carbon nexus in hybrid AC/DC microgrids for sustainable development of remote communities},
	author={Valencia-D{\'\i}az, Alejandro and Toro, Eliana M and Hincapi{\'e}, Ricardo A},
	journal={Applied Energy},
	volume={377},
	pages={124517},
	year={2025},
	publisher={Elsevier}
}

@article{liu2025optimal,
	title={Optimal {FNN}-Based Energy Management System With High Real-Time Performance and Good Interpretability for Battery in Grid-Connected Microgrid},
	author={Liu, Bin and Wang, Dan and Huang, Jiawei and Mao, Chengxiong},
	journal={IEEE Transactions on Industrial Electronics},
	year={2025},
	publisher={IEEE}
}

@article{xu2021bayesian,
	title={Bayesian adversarial multi-node bandit for optimal smart grid protection against cyber attacks},
	author={Xu, Jianyu and Liu, Bin and Mo, Huadong and Dong, Daoyi},
	journal={Automatica},
	volume={128},
	pages={109551},
	year={2021},
	publisher={Elsevier}
}

@article{tafone2025multi,
	title={Multi-objective design optimization of cryo-polygeneration systems for urban microgrids: Balancing cost-effectiveness and sustainability},
	author={Tafone, Alessio and Thangavelu, Sundar Raj and Morita, Shigenori and Romagnoli, Alessandro},
	journal={Energy},
	volume={328},
	pages={136550},
	year={2025},
	publisher={Elsevier}
}

@article{oyewole2024multi,
	title={Multi-objective optimal sizing and design of renewable and diesel-based autonomous microgrids with hydrogen storage considering economic, environmental, and social uncertainties},
	author={Oyewole, Oladimeji and Nwulu, Nnamdi and Okampo, Ewaoche John},
	journal={Renewable Energy},
	volume={231},
	pages={120987},
	year={2024},
	publisher={Elsevier}
}

@article{ghodusinejad2024multi,
	title={Multi-objective optimal design and performance analysis of a residential microgrid},
	author={Ghodusinejad, Mohammad Hasan and Peirov, Setareh and Yousefi, Hossein and Astaraei, Fatemeh Razi},
	journal={Journal of Energy Storage},
	volume={92},
	pages={112201},
	year={2024},
	publisher={Elsevier}
}

@article{rehman2025research,
	title={Research on optimal design of multi-energy microgrid considering hybrid resilience load management and Carbon emissions},
	author={Rehman, Talha and Khan, Muhammad Ahsan and Kim, Hak-Man},
	journal={Sustainable Cities and Society},
	volume={119},
	pages={106108},
	year={2025},
	publisher={Elsevier}
}

@article{macmillan2024microgrid,
	title={Microgrid design and multi-year dispatch optimization under climate-informed load and renewable resource uncertainty},
	author={Macmillan, Madeline and Zolan, Alexander and Bazilian, Morgan and Villa, Daniel L},
	journal={Applied Energy},
	volume={368},
	pages={123355},
	year={2024},
	publisher={Elsevier}
}

@article{kazemtarghi2024techno,
	title={Techno-economic microgrid design optimization considering fuel procurement cost and battery energy storage system lifetime analysis},
	author={Kazemtarghi, Abed and Mallik, Ayan},
	journal={Electric Power Systems Research},
	volume={235},
	pages={110865},
	year={2024},
	publisher={Elsevier}
}

@article{fahad2026optimal,
	title={Optimal design of a microgrid based on economic considerations for rural electrification using hybrid renewable energy system},
	author={Fahad, Shah and Khan, Shoaib Ahmed and Salman, Muhammad and Han, Yuyan and Liu, Yiping},
	journal={Energy},
	pages={140975},
	year={2026},
	publisher={Elsevier}
}

@article{bukar2025optimal,
	title={Optimal design of hydrogen storage-based microgrid employing machine learning models},
	author={Bukar, Abba Lawan and Menesy, Ahmed S and Kassas, Mahmoud and Abido, Mohammad A and Modu, Babangida and Hamza, Mukhtar Fatihu},
	journal={International Journal of Hydrogen Energy},
	volume={159},
	pages={150539},
	year={2025},
	publisher={Elsevier}
}

@article{boennec2025robust,
	title={Robust design of microgrids under uncertainties: Comparative impact of modeling choices on sizing decisions},
	author={Boennec, Corentin and Sareni, Bruno and Ngueveu, Sandra Ulrich},
	journal={Energy Conversion and Management: X},
	pages={101330},
	year={2025},
	publisher={Elsevier}
}

@article{bukar2026data,
	title={A data-driven framework for microgrid design integrating machine learning model with economic-energy-environmental parameters},
	author={Bukar, Abba Lawan and Kassas, Mahmoud and Abido, Mohammad A and Menesy, Ahmed S and Modu, Babangida and Hamza, Mukhtar Fatihu and Didane, Djamal Hissein},
	journal={Renewable Energy Focus},
	volume={56},
	pages={100785},
	year={2026},
	publisher={Elsevier}
}

@article{alshammari2026energy,
	title={How energy management strategy shapes optimal microgrid design: A comparative analysis for {EV} charging stations},
	author={Alshammari, Nahar F and Alshahr, Shahr and Barakat, Shimaa},
	journal={Applied Energy},
	volume={407},
	pages={127352},
	year={2026},
	publisher={Elsevier}
}

@article{saleheen2026ensuring,
	title={Ensuring reliability in 100\% renewable microgrids: A scenario-based joint planning and operational design framework},
	author={Saleheen, Mohammed Zeehan and Wagner, Markus and Wang, Hao},
	journal={Energy Conversion and Management},
	volume={363},
	pages={121645},
	year={2026},
	publisher={Elsevier}
}

@article{hood2025cost,
	title={Cost, resiliency and emissions trade-offs for microgrids in varying socioeconomic settings},
	author={Hood, Karoline and McMiller, Orlando and Nock, Destenie and Grymes, James and Newman, Alexandra},
	journal={Renewable and Sustainable Energy Reviews},
	volume={216},
	pages={115550},
	year={2025},
	publisher={Elsevier}
}

@article{eklund2025evaluating,
	title={Evaluating the interplay of community behaviour and microgrid design through optimisation modelling in local energy markets},
	author={Eklund, Melissa and Voinov, Alexey and Hossain, MJ and Khalilpour, Kaveh},
	journal={Renewable and Sustainable Energy Reviews},
	volume={210},
	pages={115271},
	year={2025},
	publisher={Elsevier}
}

@article{vincent2025sustainable,
	title={Sustainable microgrids design with uncertainties and blockchain-based peer-to-peer energy trading},
	author={Vincent, F Yu and Le, Thi Huynh Anh and Gupta, Jatinder ND},
	journal={Renewable and Sustainable Energy Reviews},
	volume={216},
	pages={115600},
	year={2025},
	publisher={Elsevier}
}

@article{rajendran2026meta,
	title={Meta-reinforcement learning for multi-energy building microgrids: Curriculum design, rapid adaptation, and scalable deployment},
	author={Rajendran, Sai Sasidhar Punyam and Gebremedhin, Alemayehu},
	journal={Energy Conversion and Management},
	volume={363},
	pages={121652},
	year={2026},
	publisher={Elsevier}
}

@article{ezzati2026resilient,
	title={Resilient microgrid planning for socially vulnerable communities},
	author={Ezzati, Farzane and Dong, Zhijie Sasha and Lim, Gino and Jiao, Junfeng},
	journal={Applied Energy},
	volume={410},
	pages={127434},
	year={2026},
	publisher={Elsevier}
}

@article{pradana2026unified,
	title={Unified robust--stochastic dispatch for reliable net-zero microgrids under {PV} uncertainty},
	author={Pradana, Adland and Yang, Fuwen and Lu, Junwei and Sanjari, Mohammad J},
	journal={Applied Energy},
	volume={420},
	pages={128113},
	year={2026},
	publisher={Elsevier}
}

@article{zelaschi2025effective,
	title={An effective MILP model for the optimal design of microgrids with high-reliability requirements},
	author={Zelaschi, Andrea and Pliotti, Lorenzo and Betti, Giulio and Tonno, Giovanni and Sgr{\`o}, Daniele and Martelli, Emanuele},
	journal={Applied Energy},
	volume={383},
	pages={125359},
	year={2025},
	publisher={Elsevier}
}

%

\end{document}